# Quantum-Aware Key Rank for Side-Channel Attacks evaluation

Yevgen Kotukh

**Abstract.** Side-channel attacks on block-cipher implementations rarely recover the key outright; they leave a residual key space ordered by subkey posteriors, and evaluation laboratories report the true key's position in that ordering, its key rank, against thresholds near $2^{48}$ and $2^{60}$. No evaluation framework models an adversary who collects traces classically and finishes the search on a quantum computer, although quantum search with advice provides exactly that. We give such an adversary a cost model and a metric. For a device whose key has rank $r$, we prove that any quantum algorithm with classical access to the posteriors needs $\Omega(\sqrt{r})$ cipher evaluations and that block search achieves $O(\sqrt{r})$: the per-instance advantage is exactly quadratic. The super-quadratic exponents recently reported for search with advice are averages over the prior and do not apply to any single device; we show both statements are consistent. Second, under NIST's MAXDEPTH depth budget $D$ and a Grover iteration of $g$ gates and depth $d$, the quadratic advantage holds only for ranks up to $(4D/\pi d)^2$ and collapses to a constant factor $16\, c_{\mathrm{cl}} D/(\pi^2 g d)$ beyond; with the best published AES-128 oracle and $D = 2^{40}$ the window closes near rank $2^{58.6}$. Third, we define the quantum-aware key rank, the classical rank of equal cost, computable from the rank estimators laboratories already run, and derive the classical rank a device must retain to keep a given threshold against each depth budget. Fourth, we compare the two ways of using the posteriors, block search and amplitude amplification from the product-form posterior state, and show that on Hamming-weight posteriors they are within a factor of two to four of each other, so that the cost of coherent indexing decides between them. Experiments on simulated leakage and a toy cipher confirm the predicted window: at SNR 0.125 a depth-limited quantum adversary is worth about 22 traces to a classical one, however large its gate budget.

## Introduction

This section states the problem, fixes the adversary we consider, and lists the five contributions with forward pointers to the sections that establish them.

A side-channel attack on a block-cipher implementation rarely recovers the key outright. Correlation power analysis (CPA) or template attacks return, for each subkey byte, a score or posterior distribution; the true key is then one candidate among many, and its position in the resulting ordering is its *key rank*. Evaluation laboratories treat the rank as the residual security of the device: optimal-order enumeration is practical up to roughly $2^{40}$ candidates [1], parallel and histogram-based enumeration extend this to the $2^{50}$–$2^{60}$ range depending on infrastructure [2; 3], and rank estimation bounds the rank even when enumeration is out of reach [4; 5; 6]. A commonly stated rule of thumb is that a rank below about $2^{48}$ counts as a full break and that a well-resourced adversary can enumerate up to about $2^{60}$ [7].

None of the frameworks that govern side-channel evaluation—the JIL attack-potential rating for smartcards [8], ISO/IEC 17825 [9], or the FIPS 140-3 test suite—models an adversary who, after collecting traces classically, performs the remaining enumeration on a quantum computer. The ingredients for such an adversary exist: Grover's algorithm [10], amplitude amplification from a biased initial state [11], Montanaro's quantum search with advice [12], and the explicit side-channel instantiation of Martin, Montanaro, Oswald and Shepherd [13], which obtains a quadratic reduction in the number of keys that must be tested. The only public reference to the consequence we are aware of is a remark that a classical cut-off at $2^{80}$ would not be safe post-quantum because Grover reduces it to $2^{40}$ [7].

Two recent results complicate the picture in opposite directions. On the one hand, Bashiri, Glaser, May and Nowakowski [14] and Schubert et al. [15] show that when the attacker's expected cost is averaged over the prior distribution, the quantum exponent is *super*-quadratic: the classical guesswork grows as $2^{H_{1/2}(\mathcal{D})}$ while the quantum cost grows as $2^{H_{2/3}(\mathcal{D})/2}$, and $H_{2/3} < H_{1/2}$ for every non-uniform distribution. On the other hand, the NIST post-quantum standardization process bounds every quantum attack by a maximal circuit depth MAXDEPTH $\in \{2^{40}, 2^{64}, 2^{96}\}$ [16], and Jaques, Naehrig, Roetteler and Virdia [17] show that under this bound Grover search on AES must be parallelized with a $\sqrt{\cdot}$-loss [18], so that the total gate cost becomes $G \cdot D_{\text{Grover}}/\text{MAXDEPTH}$ rather than $G$. An evaluation lab that wants a single number for "residual security against a quantum adversary" is therefore caught between a result that says the speed-up is more than quadratic and a constraint that says it may be no speed-up at all.

We resolve the tension by separating two questions that the literature has treated as one. The first is *per-instance*: a specific device holds a specific key of a specific rank $r$, and the evaluator wants the cost of recovering *that* key. The second is *average-case*: a population of devices draws keys from a distribution, and one wants the expected cost. We show that the per-instance quantum advantage is exactly quadratic in $r$ and cannot be improved, that the super-quadratic exponents are a property of the average only, and that under MAXDEPTH the quadratic advantage survives only for ranks inside an explicit window. We package the result as a metric, *quantum-aware key rank*, that an evaluation lab can compute with the rank estimators it already runs.

## Contributions.

**C1.** *Per-instance quadratic limit* (Sect. 4, Theorem 1, Corollary 1). For a key of residual rank $r$, any quantum algorithm with classical access to the posteriors and quantum access to the cipher oracle needs $\Omega(\sqrt{r})$ queries, and block search achieves $O(\sqrt{r})$. The super-quadratic exponents of [14; 15] arise only under averaging over the prior; conditioning on the rank removes them. Both statements are correct and answer different questions.

**C2.** *Quantum window* (Sect. 5, Theorem 3). With a Grover iteration of $g$ gates and depth $d$ and a depth budget $D$, the quantum cost is $\Theta(\sqrt{r})$ only for $r \leq r_{\text{hi}} = (4D/\pi d)^2$; above it the cost is $\Theta(r)$ and the gain over classical enumeration is the constant $16/\pi^2\, c_{\text{cl}} D/(gd)$. We give closed-form window boundaries and their sensitivity to the published value of $d$.

**C3.** *Quantum-aware key rank* (Sect. 3, Definitions 1–3; Sect. 4, Proposition 5). A family of metrics—per-instance QR, depth-constrained $\text{QKR}_D$ defined as the classical rank of equal cost, and average-case quantum guesswork QG—computable from histogram and ESrank estimators.

**C4.** *Trace discount* (Sect. 7, Proposition 6). For a fixed gate budget, how many additional traces a classical attacker needs to match a quantum attacker, predicted in closed form and measured on simulated Hamming-weight leakage: the discount plateaus at $\log_2 \gamma/\mu$ traces, about 22 at $\text{SNR} = 0.125$ and $D = 2^{40}$.

**C5.** *Which quantum strategy wins* (Sect. 6, Proposition 4). Amplitude amplification from a cheap product-form initial state costs $\approx 1/\sqrt{p_t}$; block search costs $\approx \sqrt{r}$ but needs coherent indexing. We prove $\sqrt{r} \leq 1/\sqrt{p_t}$ always, give the condition on the advice sharpness $\sigma = 1/\sqrt{r p_t}$ and the indexing overhead $\kappa$ under which block search wins in gates, and measure $\sigma$ on 4 000 simulated instances: it is 2–3 bits, so the two strategies are within a small constant of each other and $\kappa$ decides.

We treat the cipher oracle's cost $(g, d)$ as an input taken from published AES circuits [17; 19; 20; 21] and report every window with its sensitivity to the choice. New circuit designs, the concrete gate cost of the coherent indexing circuit, and the Ukrainian standard Kalyna are the subject of a companion paper [22] and appear here only as parameters or as future work.

Section 2 reviews side-channel posteriors, classical enumeration, Grover-type search, search with advice, and the MAXDEPTH cost model, and identifies the gap. Section 3 fixes the adversary, the notation and the metrics. Sections 4–6 contain the theory (C1, C2, C5). Section 7 reports experiments (C4, C5). Section 8 draws the consequences for evaluation practice, and Sect. 9 states limitations. Proofs are in Appendix A.

With the problem and the contributions in place, we next collect the background each contribution builds on.

## Background

This section reviews the five bodies of work the paper combines and closes each with what that work does not do.

A divide-and-conquer side-channel attack on a block cipher targets an intermediate value that depends on the plaintext and on one subkey byte at a time, typically the S-box output of the first round. For $q$ traces, CPA [23] computes, for each hypothesis $j$ of subkey $i$, the Pearson correlation between the measured leakage and a leakage model of the intermediate value (Hamming weight in the simplest case); a template attack [24] instead fits a Gaussian model per intermediate value during a profiling phase and returns, for each hypothesis, a likelihood. Either way the attack ends with an $m \times n$ array of scores, and under the usual Bayesian treatment with a uniform prior on keys these scores become posteriors $p_{i,j} = \Pr[k_i = j \mid \text{leakage}]$ with $\sum_j p_{i,j} = 1$ for each $i$. The posterior on the full key is taken to be the product $\prod_i p_{i,k_i}$; this independence assumption is standard in the enumeration literature and is exact when the leakages of different subkeys are conditionally independent given the key, which holds for unprotected first-round S-box leakage and fails, for instance, for masked implementations where the mask ties several bytes together. The side-channel literature establishes how posteriors are obtained and how good they are; it does not consider what an adversary with a quantum computer does with them.

Given the posteriors, two classical problems arise. *Enumeration* outputs keys in decreasing order of probability until the true key is found; the optimal-order algorithm of Veyrat-Charvillon, Gérard, Renauld and Standaert [1] is memory-bound and practical to about $2^{40}$ candidates. *Rank estimation* does not output keys but bounds the position of a known key in the ordering; it was introduced in [4] and made fast and tight by the histogram-convolution method of Glowacz et al. [5], the path-counting method of Martin, O'Connell, Oswald and Stam [2] (which also enumerates in parallel), the histogram enumeration of Poussier, Standaert and Grosso [3], the bounded-space enumeration of David and Wool [25], and ESrank [6], which estimates the rank to within about one bit in under a second; Martin, Mather and Oswald [26] show that path counting and histogram convolution are two views of the same computation. Bogdanov et al. [27] give a score-based enumeration with low memory, and Villanueva-Polanco [28] surveys the problem. All of these algorithms work on integer weights obtained by discretizing $-\log p_{i,j}$ and add keys by weight; a key's rank is then the number of keys of strictly smaller weight, up to a discretization error that is analyzed in [13]. The practical conclusion drawn by the community is that ranks up to roughly $2^{50}$–$2^{60}$ are reachable by a determined classical adversary and that an evaluator therefore reports $\log_2 \text{rank}$ with these thresholds in mind [7]. This literature determines the classical cost of the residual search exactly; it does not model a quantum adversary, and its thresholds are stated for classical enumeration only.

Grover's algorithm [10] finds the unique marked item among $N$ with $\lfloor \frac{\pi}{4}\sqrt{N} \rfloor$ applications of the oracle, and Boyer, Brassard, Høyer and Tapp [29] give the success probability after $j$ iterations as $\sin^2\big((2j+1)\theta\big)$ with $\sin^2\theta = 1/N$, together with a strategy for an unknown number of marked items. Amplitude amplification [11] generalizes the algorithm to an arbitrary initial state $A|0\rangle$: if the marked item has amplitude $\sqrt{a}$ in $A|0\rangle$, it is found after $O\big(1/\sqrt{a}\big)$ iterations of the operator $AS_0A^{-1}S_f$. When $a$ is unknown, the fixed-point variant of Yoder, Low and Chuang [30] reaches any target success probability with an optimal number of queries without overshooting. Two limits govern everything that follows. First, $\Omega\big(\sqrt{N}\big)$ queries are necessary [31], and Zalka [18] proves that Grover's algorithm is exactly optimal: for every number of queries up to $\approx \frac{\pi}{4}\sqrt{N}$ it attains the largest success probability any quantum algorithm can. Second, Zalka also shows that running $S$ Grover instances in parallel reduces the number of iterations each must perform only by $\sqrt{S}$, so the total number of oracle evaluations grows by $\sqrt{S}$; parallelization is possible but not free. Grover-type search fixes the query complexity of unstructured search; it says nothing about how to exploit a prior that is known classically.

Montanaro [12] considers search over $N$ items when the algorithm is given, as advice, a probability distribution over the location of the marked item. Sorting the items by decreasing probability and running Grover on blocks of geometrically increasing size gives an expected number of queries $O\big(\sum_i p_{(i)}\sqrt{i}\big) = \Theta\big(\mathbb{E}\big[\sqrt{G}\big]\big)$, where $G$ is the (random) position of the marked item in the sorted order; Montanaro proves this is optimal up to a constant for any fixed distribution. The algorithm assumes the advice is presented already sorted. Martin, Montanaro, Oswald and Shepherd [13] remove this assumption for the side-channel setting: using the path-counting structure of [2] they construct a function getKey$(b, w, W_1, W_2, r)$ that returns the $r$-th key of weight in $[W_1, W_2)$ in time $O(m^2 n \log n)$, so that Grover can search *by index* inside each weight slice without ever materializing the sorted list; the result is a quadratic reduction in the number of keys that must be tested, at the price of a quantum random-access memory (QRAM) holding the path-counting array and of a coherent evaluation of getKey inside every oracle call. Banegas and Villanueva-Polanco [32] instantiate the same idea for cold-boot key recovery and provide Grover oracles for AES, PRESENT, GIFT and LowMC; Albrecht and Shen [33] use search with advice and a bounded-error oracle [34] inside a dual attack on LWE.

The exact exponent of the average-case speed-up was settled recently. Bashiri, Glaser, May and Nowakowski [14] prove that the cost of Montanaro's algorithm on a product distribution $\mathcal{D}$ is $2^{H_{2/3}(\mathcal{D})/2}$ while classical guessing costs $2^{H_{1/2}(\mathcal{D})}$ by Arıkan's bound [35], so the exponent ratio exceeds 2 for every non-uniform $\mathcal{D}$; Schubert, Paskarbeit, Kramer, Seifert and Margraf [15] compute this ratio exactly or with a certified error for cold-boot, template and Keccak side-channel distributions and observe ratios up to about 3.97 on synthetic models. Both papers measure expected query complexity under the prior; neither accounts for the gate cost of the cipher oracle, for the cost of coherent indexing, or for a depth bound, and neither asks what the cost is for a device whose key has a given rank. This paper takes up exactly those questions, and Sect. 4 shows how the two viewpoints are reconciled.

## Gap analysis

To our knowledge, no published work (i) states the per-instance quantum cost as a function of the rank, (ii) combines search with advice with the MAXDEPTH cost model and concrete oracle costs, (iii) quantifies the trade-off between side-channel traces and quantum budget, or (iv) proposes a metric that an evaluation lab could report. Related topics that are *not* this problem include quantum attacks on QKD hardware (also called "quantum side channels"), fault attacks whose residual space is small enough that no quantum speed-up matters, and side channels on lattice-based schemes [15]; we position against them in Sect. 9 only.

## Quantum Resource Estimates for AES and the MAXDEPTH Model

NIST's call for post-quantum proposals [16] defines security categories by the cost of Grover key search on AES and introduces MAXDEPTH $\in \{2^{40}, 2^{64}, 2^{96}\}$ as the maximal depth of any quantum attack circuit, justified as the number of sequential gates a quantum computer could perform in roughly a year, a decade, and a millennium respectively. A Grover search whose unparallelized depth $D_{\text{Grover}}$ exceeds MAXDEPTH must be split into $x^2$ parallel instances each running for $1/x$ of the time, so that the total gate count becomes $G \cdot D_{\text{Grover}}/\text{MAXDEPTH}$; NIST's table gives $2^{170}/\text{MAXDEPTH}$ quantum gates for AES-128 against $2^{143}$ classical gates. Jaques, Naehrig, Roetteler and Virdia [17] implement the full Grover oracle for AES in Q#, derive the parallelization formulas we use in Sect. 5 (their Eqs. (5)–(9)), and lower the AES-128 estimate to about $2^{157}/\text{MAXDEPTH}$; they also show that for MAXDEPTH $= 2^{96}$ no parallelization is needed for AES-128. Their AES-128 oracle with one plaintext–ciphertext pair and Maximov's MixColumns has full depth 2086 and width 2817, with 294 863 CNOT, 84 488 one-qubit Clifford, 54 908 $T$ and 13 727 measurement gates ([17, Table 8], arXiv version; the EUROCRYPT version may differ in the last digits [VERIFY]). Subsequent work lowered the depth further: Huang and Sun [19] reduce the $T$-depth of AES-128 from 60 to 40 at equal width or to 30 with a small width increase, and reduce the width of the narrowest circuit from 512 [36] to 371; Liu, Preneel, Zhao and Wang [20] [VERIFY] reduce the full depth of the AES-128 encryption circuit to 730 in a pipelined architecture, and Jiang et al. [21] tabulate the full pipeline (compute-and-uncompute) circuits of these works side by side, which is the form our cost model needs; Jang et al. [37] report a pool of 78 implementations and an attack cost of $2^{156.26}$ for AES-128 in NIST's accounting. The earlier estimates of Grassl et al. [38], which NIST's original table is based on, and of Bonnetain, Naya-Plasencia and Schrottenloher [39], who show that no structural attack beats generic search on full AES by a meaningful margin, complete the picture. This literature supplies the per-iteration gate count $g$ and depth $d$ that our cost model needs and the parallelization rule that turns a depth bound into a gate count; it does not consider a search space smaller than $2^k$ or a non-uniform prior, and so it does not say at what rank the depth bound starts to matter.

Having collected the ingredients, we now fix the adversary model, the notation, and the metrics that the rest of the paper uses.

# Model, Notation and Metrics

This section defines the adversary, lists every symbol used in the paper, and introduces the three quantum-aware rank quantities.

## Adversary Model

We consider a Q1 adversary in the sense of [40]: all interaction with the target device is classical, and quantum computation happens offline. Concretely, the adversary (i) collects $q$ power or electromagnetic traces together with the corresponding plaintexts and at least one plaintext–ciphertext pair $(x_0, c_0)$; (ii) runs a classical CPA or template attack to obtain posteriors $\{p_{i,j}\}$; (iii) builds, from the known pair, a quantum circuit $O_{k^*}$ that marks a candidate key $k$ iff $E_k(x_0) = c_0$; and (iv) runs a quantum algorithm with classical access to $\{p_{i,j}\}$ and quantum access to $O_{k^*}$. The adversary has no quantum access to the device. The total depth of the quantum computation is bounded by $D \in \{2^{40}, 2^{64}, 2^{96}\}$, and we charge the number of logical gates; error-correction overhead is outside the model (Sect. 9). Two assumptions are used throughout.

**Assumption 1 (Independence).** The posterior on the full key factors: $\Pr[k = (j_1, \dots, j_m) \mid \text{leakage}] = \prod_{i=1}^{m} p_{i,j_i}$.

**Assumption 2 (Unique marked key).** Among the keys of rank at most $r$ there is exactly one key $k$ with $E_k(x_0) = c_0$, namely $k^*$.

Assumption 1 is the standard enumeration model of Sect. 2.2. Assumption 2 holds with probability $\geq 1 - r \cdot 2^{-n_b}$ for a block size $n_b$ under the usual random-permutation heuristic [17, Sect. 2.2], which is overwhelming for $r \leq 2^{100}$ and $n_b = 128$; a second pair removes the issue entirely at twice the oracle cost.

## Per-iteration cost.

One iteration of Grover's algorithm or of amplitude amplification consists of one call to $O_{k^*}$ and one reflection; for block search the reflection is about the uniform state on the current block, for product-state amplification about the product state. The cipher oracle dominates both, so we take $g$ and $d$ to be the oracle's gate count and depth plus the reflection's and treat $\kappa$ as the only quantity that distinguishes the two strategies (Sect. 6). For AES-128 we use five published circuits plus the Grassl-era anchor and report every result for each; Table 2 lists them.

***Table 2.*** *Per-iteration costs for the AES-128 oracle used as inputs. For the three circuits after [17] the tabulated values are those of the encryption circuit as reported in [21, Table 8] (Liu et al. there doubled from a forward-only circuit), and $g$ adds one-qubit Clifford and measurement gates at the JNRV20 ratio*

$0.332 \times \#CNOT$ *(estimate). The takeaway is that* $d$ *varies by a factor of about* $1.8$ *across post-2020 circuits and by* $90\times$ *against the Grassl-era circuit behind NIST's figure.*

| Source | circuit | $g$ (gates) | $d$ (full depth) | width | note |
|---|---|---|---|---|---|
| [38] via [17, Tab. 10] | Grover oracle, $r = 3$ | $1.46 \cdot 10^7$ | 193 588 | 2953 | G-cost, depth ÷ $\lfloor \pi/4\, 2^{64} \rfloor$ (derived) |
| [17] Tab. 8 | oracle, in-place MC, $r = 1$ | 445 376 | 2816 | 1665 | CNOT+1qC+T+M |
| [17] Tab. 8 | oracle, Maximov MC, $r = 1$ | 447 986 | 2086 | 2817 | CNOT+1qC+T+M |
| [19] | encryption, pipeline | 452 340 | 2198 | 3936 | 298 720 CNOT, 54 400 $T$ (est.) |
| [20] | encryption, pipeline (× 2) | 407 050 | 1600 | 3689 | 264 752 CNOT, 54 400 $T$ (est.) |
| [21] | encryption, pipeline | 306 548 | 1790 | 2896 | 189 300 CNOT, 54 400 $T$ (est.) |

Two numbers recur. From the NIST table [16], AES-128 costs $2^{170}/\text{MAXDEPTH}$ quantum gates and $2^{143}$ classical gates over $2^{128}$ keys; since the parallelized cost equals $(\pi/4)^2\, 2^{128}\, gd/\text{MAXDEPTH}$ (Sect. 5), this gives $gd \approx 2^{42}$ in NIST's accounting and $c_{\text{cl}} \approx 2^{143-128} = 2^{15}$ per key *(both derived)*. With the Jaques et al. oracle (Maximov MC, $r = 1$) one has $gd = 447\,986 \times 2086 \approx 1.74 \cdot 2^{29}$ *(derived)*, and indeed $(\pi/4)^2 \cdot 2^{128} \cdot 1.74 \cdot 2^{29} \approx 1.07 \cdot 2^{157}$ reproduces their Table 12 *(derived check)*. We carry both anchors, $gd \approx 2^{42}$ [16] and $gd \approx 2^{29.8}$ ([17]), since the window boundaries depend on them.

## Metrics

We now define the three quantum-aware quantities. The first is the per-instance analogue of the key rank.

**Definition 1 (Per-instance quantum rank).** $\text{QR}(k^*) := \left\lceil \sqrt{\text{rank}(k^*)} \right\rceil$.

QR is the number of oracle queries, up to a constant, that an unconstrained quantum adversary needs (Theorem 1). It ignores gate cost and depth; the next definition restores both and expresses the result in the unit evaluation labs already use.

**Definition 2 (Depth-constrained quantum-aware key rank).** For a depth budget $D$, per-iteration cost $(g, d)$ and classical per-key cost $c_{\text{cl}}$,

$$\text{QKR}_D(k^*) \;:=\; \frac{C_q^D\big(\text{rank}(k^*)\big)}{c_{\text{cl}}},$$

the classical rank whose optimal-order enumeration costs as many gates as the depth-limited quantum attack on $k^*$.

$\text{QKR}_D$ is dimensionless, monotone in the rank, and directly comparable with thresholds such as $2^{48}$ or $2^{60}$: a device with $\text{QKR}_{2^{40}}(k^*) = 2^{45}$ is, against a quantum adversary limited to depth $2^{40}$, exactly as secure as a device of classical rank $2^{45}$ against a classical one. Section 5 gives $\text{QKR}_D$ in closed form (Corollary 2). The third quantity is the average-case counterpart.

**Definition 3 (Quantum guesswork).** $\mathrm{QG}(\mathcal{D}) := \mathbb{E}_{K\sim\mathcal{D}}\left[\sqrt{\mathrm{rank}(K)}\right] = \mathbb{E}[\sqrt{G}]$.

QG is the expected query cost of Montanaro's algorithm up to a constant [12] and equals $2^{H_{2/3}(\mathcal{D})/2(1+o(1))}$ for product priors [14]; it is the quantity whose exponent is super-quadratic relative to the classical guesswork $\mathbb{E}[G]$. Proposition 5 in Sect. 4 shows that QR and $\mathrm{QKR}_D$ inherit the bounds of any rank estimator, so an evaluator can report them without new tooling.

With the model and the metrics fixed, Sect. 4 proves that the per-instance speed-up is exactly quadratic and explains why the average-case exponent is larger.

This section proves that the quantum query advantage for a key of given rank is exactly quadratic, restates the average-case result of [14] in our notation, and shows that the two statements are consistent.

## The Black-Box Model and Rank Classes

Throughout this section the oracle is the idealized indicator $O_{k^*}: |k\rangle|y\rangle \mapsto |k\rangle|y \oplus [k = k^*]\rangle$; by Assumption 2 and the discussion following it, this coincides with the cipher oracle except with negligible probability, and a second plaintext–ciphertext pair makes the two identical at twice the cost. The posteriors $\{p_{i,j}\}$, and hence the ordering of $\mathcal{K}$ by $P$, are fixed and known to the algorithm; only $k^*$ varies. A statement "for a key of rank $r$" must be read with care: the posteriors determine which key has rank exactly $r$, so an algorithm tailored to a single rank could output that key with one query. The meaningful per-instance notion is therefore the *rank class*

$$B_r := \{k \in \mathcal{K}:\ r/2 < \mathrm{rank}(k) \leq r\}, \qquad |B_r| = \lfloor r/2 \rfloor,$$

and we ask for the cost of an algorithm that must work for every key in the class, equivalently for an algorithm that does not know the rank in advance.

**Theorem 1 (Per-instance quadratic limit).** Fix the posteriors and let $r \geq 8$.

1. (a) *(Lower bound.)* Let $\mathcal{A}$ be a quantum algorithm with classical access to $\{p_{i,j}\}$ and query access to $O_{k^*}$ that outputs $k^*$ with probability at least $p$ for every $k^* \in B_r$. Then $\mathcal{A}$ makes at least $c_p\sqrt{r/2}\,(1 - o(1))$ queries on some $k^* \in B_r$, where $c_p = \arcsin(\sqrt{p})/2$; the same bound holds for the expected number of queries averaged over $k^*$ uniform in $B_r$. In particular $\Omega(\sqrt{r})$ queries are necessary.

2. (b) *(Upper bound.)* Block search (Algorithm 1) does not know the rank, outputs $k^*$ with probability at least $3/4$ for every $k^* \in \mathcal{K}$, and makes at most

   $$\pi/4 \cdot \sqrt{2}/\sqrt{2} - 1\,\sqrt{\mathrm{rank}(k^*)} + O(\log \mathrm{rank}(k^*)) \ < \ 2.69\sqrt{\mathrm{rank}(k^*)} + O(\log \mathrm{rank}(k^*))$$

   queries.

Consequently, for every rank class the optimal query count is $\Theta(\sqrt{r})$, and $\log_2 \mathrm{QR}(k^*) = 1/2\log_2 \mathrm{rank}(k^*) + O(1)$ is the per-instance cost of an optimal quantum adversary up to an additive constant.

**Algorithm 1** (Block search with unknown rank (after [12; 13], block ratio 2))

1. Test classically the 7 keys of rank $\leq 7$; if one is $k^*$, return it.

2. For $s = 3,4,5, \ldots$:

3. Let $R_s := \{k\colon\ 2^s \le \mathrm{rank}(k) < 2^{s+1}\}$, $|R_s| = 2^s$.
4. Prepare the uniform superposition over $R_s$ via coherent indexing $\iota_s\colon [2^s] \to R_s$ (Sect. 6).
5. Run $\lfloor \pi/4\, 2^{s/2} \rfloor$ Grover iterations with oracle $O_{k^*} \circ \iota_s$ and measure a candidate $\hat{k}$.
6. Test $\hat{k}$ classically; if $\hat{k} = k^*$ return it.
7. End for.

The proof is in Appendix A.1. Three remarks on what the theorem does and does not say.

**Remark 1 (Why the constant in (a) is exact).** Zalka [18] shows that for any number of queries $j$, Grover's algorithm on $M$ items attains the largest success probability any algorithm can, namely $\sin^2\big((2j+1)\theta\big)$ with $\sin\theta = 1/\sqrt{M}$. Inverting this at $M = |B_r|$ gives the constant $c_p$; for $p \to 1$, $c_p \to \pi/4$, the same constant that appears in the cost model of Sect. 5. The lower bound therefore matches the upper bound of (b) up to the geometric factor $\sqrt{2}/\big(\sqrt{2}-1\big) \approx 3.41$ incurred by not knowing the rank, which Montanaro [12] shows can be reduced but not removed.

**Remark 2 (The advice is used only to order).** In (a), all keys in $B_r$ share the same posteriors, so the advice cannot distinguish among them: once the ordering is fixed, the residual problem is unstructured search on $|B_r|$ items. This is the formal content of the claim that the per-instance advantage is "exactly quadratic": the prior can shrink the search space to the top-$r$ keys, but inside the rank class nothing better than Grover is possible.

**Remark 3 (Oracle versus gate cost).** Theorem 1 counts queries. The coherent indexing $\iota_s$ in Algorithm 1 is free in the query model but not in gates; its cost is the parameter $\kappa$ of Sect. 6.

The average-case cost is a statement about the prior $\mathcal{D}$, not about a key. We restate it in our notation; the proof is a direct reading of [35; 12; 14] and is given in Appendix A.2 for completeness.

**Proposition 1 (Average-case exponents).** Let $\mathcal{D} = \mathcal{D}_1 \otimes \cdots \otimes \mathcal{D}_m$ be a product prior on $[n]^m$ and $G = \mathrm{rank}(K)$ for $K \sim \mathcal{D}$. Then, with $\alpha$-Rényi entropies additive over the product,

$$\begin{aligned} \log_2 \mathbb{E}[G] &= H_{1/2}(\mathcal{D}) + O(\log m), \\ \log_2 \mathbb{E}\big[\sqrt{G}\big] &= 1/2\, H_{2/3}(\mathcal{D}) + O(\log m). \end{aligned}$$

The optimal expected quantum query cost with advice $\mathcal{D}$ is $\Theta\big(\mathbb{E}\big[\sqrt{G}\big]\big)$ [12]. Hence the exponent ratio $s := \log\mathbb{E}[G]/\log\mathbb{E}\big[\sqrt{G}\big] \to 2H_{1/2}(\mathcal{D})/H_{2/3}(\mathcal{D}) \ge 2$, with equality if and only if $\mathcal{D}$ is uniform.

**Corollary 1 (Reconciliation).** Proposition 1 and Theorem 1 are consistent. Conditioning on the rank class $B_r$ removes the average over $\mathcal{D}$ that produces $s > 2$: within each class the exponent is exactly $1/2$. The super-quadratic average arises because the quantum cost $\sqrt{G}$ is concave, so that the mass $\mathcal{D}$ places on small ranks weighs more in $\mathbb{E}\big[\sqrt{G}\big]$ than in $\mathbb{E}[G]$ relative to the mass on large ranks. For the evaluation of a specific device, whose key and rank are fixed, Theorem 1 is the relevant statement; for the security of a key-generation process across a population of devices, Proposition 1 is.

**Example 1 (Worked example).** Take posteriors with $m = 16$ bytes in which each byte has one value of probability $1/2$ and 255 values of probability $1/510$. Per byte, $H_{1/2} = 2\log_2\big(\sqrt{1/2} + 255\sqrt{1/510}\big) =$ 7.17 bits and $H_{2/3} = 3\log_2\big((1/2)^{2/3} + 255\,(1/510)^{2/3}\big) = 6.63$ bits *(derived)*, so $H_{1/2}(\mathcal{D}) = 114.7$ and $H_{2/3}(\mathcal{D}) = 106.0$ bits for the full key. Classical guesswork is about $2^{114.7}$, quantum guesswork about $2^{53.0}$, and the average-case exponent ratio is $s = 2 \times 7.17/6.63 = 2.16$. A device whose key happens to have rank $2^{56}$ under these posteriors is nevertheless attacked with $\Theta(2^{28})$ queries, and one whose key has rank $2^{100}$ with $\Theta(2^{50})$: the ratio $s$ describes the ensemble, not either device.

**Proposition 2 (Estimator compatibility).** Let a rank estimator return $r_\ell \le \widehat{\mathrm{rank}}(k^*) \le r_u$ on the weight-discretized problem, and let the discretization distort ranks by at most a factor $\rho \ge 1$, i.e. $\mathrm{rank}(k)/\rho \le \widehat{\mathrm{rank}}(k) \le \rho\, \mathrm{rank}(k)$ for all $k$. Then

$$\sqrt{r_\ell/\rho} \;\le\; \mathrm{QR}(k^*) \;\le\; \sqrt{\rho\, r_u}\,,$$

and for every $D$, $\mathrm{QKR}_D$ is non-decreasing in the rank, so

$$\mathrm{QKR}_D(k^*) \in \left[\, C_q^D(r_\ell/\rho)/c_{\mathrm{cl}},\; C_q^D(\rho r_u)/c_{\mathrm{cl}} \,\right].$$

The quantity $\mathrm{QG}(\mathcal{D})$ requires no estimator: with Assumption 1 the exponent in (2) is $1/2 \sum_i H_{2/3}(\mathcal{D}_i)$, computable in $O(mn)$ time from the posteriors.

The factor $\rho$ depends on the number of bits kept when rounding $-\log p_{i,j}$ to integers; Martin et al. [13, Sect. 3] analyze the resulting ordering error; with $2048$ bins per byte the histogram bounds measured in Sect. 7 differ by a median of $0.009$ bits and by at most $1.6$ bits, so $\log_2 \rho < 1$ for all ranks above $2^4$. An evaluator who already reports a rank interval can therefore report QR and $\mathrm{QKR}_D$ intervals from the same run.

Theorem 1 fixes the query cost; the next section converts queries into gates under a depth bound, which is where the quadratic advantage acquires an upper limit.

This section derives the gate cost of a depth-limited quantum adversary as a function of the rank, identifies the rank window in which the advantage is quadratic, and gives $\mathrm{QKR}_D$ in closed form.

We follow the parallelization accounting of Jaques et al. (Jaques et al. 2020, Sect. 3.4, Eqs. (5)–(9)), with the search space of size $2^k$ replaced by the rank class of size $r$ and with inner parallelization (splitting the block into $S$ disjoint parts, each searched by one machine). Let one iteration cost $g$ gates and depth $d$, let the target success probability be $p$ with $c_p = \arcsin(\sqrt{p})/2$ ($c_1 = \pi/4$), and let $I_p(r) = \lceil c_p \sqrt{r} \rceil$.

**Lemma 1 (Two regimes).** Under depth budget $D$, the total gate cost of finding a key of rank $r$ with success probability $p$ is

$$C_q^D(r) \;=\; \begin{cases} I_p(r)\, g \;\approx\; c_p \sqrt{r}\, g & \text{if } I_p(r)\, d \le D \text{ (Regime A)}, \\ S \cdot \lfloor D/d \rfloor\, g \;\approx\; c_p^2 \dfrac{r\, g\, d}{D} & \text{if } I_p(r)\, d > D \text{ (Regime B)}. \end{cases}$$

In Regime B, $S \approx c_p^2\, r\, d^2/D^2$ machines run in parallel. The boundary between the regimes is $r_{\mathrm{hi}}(D, d) := (D/(c_p\, d))^2$, and at the boundary both expressions equal $gD/d$.

The proof (Appendix A.3) is the computation of [17] with $2^k$ replaced by $r$; the $\sqrt{S}$ loss is Zalka's [18]. The regime-B cost is *linear* in $r$: once the depth budget forces parallelization, every doubling of the rank doubles the number of machines and the gate count, exactly as in classical enumeration.

**Theorem 2 (Quantum window).** Let $C_{\mathrm{cl}}(r) = c_{\mathrm{cl}}\, r$ and define

$$r_{\mathrm{lo}} := \left(\frac{c_p\, g}{c_{\mathrm{cl}}}\right)^2, \qquad r_{\mathrm{hi}}(D, d) := \left(\frac{D}{c_p\, d}\right)^2, \qquad \gamma(D) := \frac{c_{\mathrm{cl}}\, D}{c_p^2\, g\, d}.$$

1. (i) For $r \le r_{\text{hi}}$ the quantum cost is $\Theta(\sqrt{r})$, and $C_q^D(r) < C_{\text{cl}}(r)$ if and only if $r > r_{\text{lo}}$.
2. (ii) For $r > r_{\text{hi}}$ the quantum cost is $\Theta(r)$, and $C_{\text{cl}}(r)/C_q^D(r) = \gamma(D)$, a constant independent of $r$.
3. (iii) The window $(r_{\text{lo}}, r_{\text{hi}}]$ in which the advantage is quadratic is non-empty if and only if $\gamma(D) > 1$, i.e. $D > c_p^2\, gd/c_{\text{cl}}$, and its width in bits is $\log_2(r_{\text{hi}}/r_{\text{lo}}) = 2\log_2\gamma(D)$.

Part (iii) is the statement an evaluator needs: the quadratic regime spans exactly twice as many bits as the logarithm of the constant-factor gain that remains outside it, and both are governed by the single dimensionless ratio $\gamma(D) = c_{\text{cl}}D/(c_p^2 gd)$. Proof in Appendix A.4.

**Corollary 2 (QKR$_D$ in closed form).** With $p \to 1$ ($c_p = \pi/4$),

$$\text{QKR}_D(k^*) \;\approx\; \begin{cases} \dfrac{\pi}{4}\dfrac{g}{c_{\text{cl}}}\sqrt{r} & r \le r_{\text{hi}}(D,d), \\ \dfrac{r}{\gamma(D)} & r > r_{\text{hi}}(D,d), \end{cases}$$

equivalently $\log_2 \text{QKR}_D = 1/2\,\log_2 r + \log_2(\pi g/4c_{\text{cl}})$ below $r_{\text{hi}}$ and $\log_2 r - \log_2\gamma(D)$ above it. The two branches agree at $r = r_{\text{hi}}$. The inverse map, which turns a classical threshold $r_{\text{th}}$ into the classical rank a device must have to retain the same security against the depth-limited quantum adversary, is

$$r_{\text{th}}^{(q)}(D) = \min\{(4c_{\text{cl}}/\pi g\; r_{\text{th}})^2,\ \gamma(D)\, r_{\text{th}}\}\ \text{ restricted to the branch in which it lies.}$$

Table 3 evaluates Theorem 2 for the per-iteration costs of Table 2, with $c_{\text{cl}} = 2^{15}$ (derived from NIST's $2^{143}/2^{128}$, Sect. 3.2) and $c_p = \pi/4$. The row labelled GLRS16 uses the per-iteration cost implied by Grassl et al.'s circuit as reproduced in [17, Table 10]: $g = 1.37 \cdot 2^{87}/\lfloor \pi/4\, 2^{64} \rfloor \approx 2^{23.80}$ and $d = 1.16 \cdot 2^{81}/\lfloor \pi/4\, 2^{64} \rfloor \approx 2^{17.56}$ *(derived)*; this is the circuit behind NIST's $2^{170}$/MAXDEPTH figure and therefore the anchor a conservative evaluator would use. All logarithms are base 2 and all entries are derived from the formulas above; the arithmetic is reproduced in Appendix A.6.

***Table 3.*** *Quantum window for AES-128 as a function of the oracle circuit and $MAXDEPTH$, $c_{cl} = 2^{15}$. The takeaway is that all post-2020 circuits agree on $r_{hi}$ to within 1.6 bits and on $\gamma$ to within 1.2 bits at every $MAXDEPTH$, while the Grassl-era circuit behind NIST's $2^{170}$ figure differs by 13 bits; an evaluation rule must therefore name its circuit generation, not its circuit. Entries > 128 bits mean the regime boundary lies beyond the full key space.*

| Oracle | $\log g$ | $\log d$ | $\log gd$ | $\log r_{\text{lo}}$ | $\log r_{\text{hi}}$ for $D =$ | | | $\log\gamma(D)$ for $D =$ | | |
|---|---|---|---|---|---|---|---|---|---|---|
| | | | | | $2^{40}$ | $2^{64}$ | $2^{96}$ | $2^{40}$ | $2^{64}$ | $2^{96}$ |
| GLRS16 (via [17] Tab. 10), $r = 3$ | 23.80 | 17.56 | 41.37 | 16.9 | 45.6 | 93.6 | 157.6 | 14.3 | 38.3 | 70.3 |
| JNRV20 in-place MC, $r = 1$ | 18.76 | 11.46 | 30.22 | 6.8 | 57.8 | 105.8 | 169.8 | 25.5 | 49.5 | 81.5 |
| JNRV20 Maximov MC, $r = 1$ | 18.77 | 11.03 | 29.80 | 6.8 | 58.6 | 106.6 | 170.6 | 25.9 | 49.9 | 81.9 |
| HS22 pipeline ($g$ est.) | 18.79 | 11.10 | 29.89 | 6.9 | 58.5 | 106.5 | 170.5 | 25.8 | 49.8 | 81.8 |
| LPZW23 pipeline ($g$ est.) | 18.63 | 10.64 | 29.28 | 6.6 | 59.4 | 107.4 | 171.4 | 26.4 | 50.4 | 82.4 |
| JCGQJW25 pipeline ($g$ est.) | 18.23 | 10.81 | 29.03 | 5.8 | 59.1 | 107.1 | 171.1 | 26.7 | 50.7 | 82.7 |

Three observations follow from the table and will be used in Sect. 8.

*The lower boundary is irrelevant in practice.* $r_{\text{lo}}$ is below $2^{17}$ even for the deepest circuit, so wherever enumeration is non-trivial a depth-unconstrained quantum adversary pays fewer logical gates than a classical one. (Error-correction overhead, which the model excludes, is what makes the comparison realistic; Sect. 9.)
*At* $\text{MAXDEPTH} = 2^{40}$ *the upper boundary sits in the range that matters.* With every post-2020 oracle the quadratic regime ends between $2^{57.8}$ and $2^{59.4}$, with the Grassl-era oracle near $2^{46}$. The classical thresholds $2^{48}$ and $2^{60}$ of Sect. 1 fall inside or at the edge of this range, so the choice of circuit decides whether a borderline device is attacked in Regime A or B.
*At* $\text{MAXDEPTH} \geq 2^{64}$ *the window covers every rank an evaluator will encounter.* The boundary exceeds $2^{105}$ for all modern circuits, so the per-instance cost is simply $\pi/4\ (g/c_{\text{cl}})\sqrt{r}$ throughout.

**Example 2 (Reading the table as an evaluator).** For the JNRV20 Maximov oracle, $\log_2(\pi g/4c_{\text{cl}}) = 3.42$. A device of classical rank $2^{56}$ attacked under $D = 2^{40}$ lies in Regime A, and $\log_2 \text{QKR}_{2^{40}} = 28 + 3.4 = 31.4$: the device is as secure as one of classical rank $2^{31}$. The same device under the Grassl-era anchor lies in Regime B and has $\log_2 \text{QKR}_{2^{40}} = 56 - 14.3 = 41.7$. Either way it is below the $2^{48}$ threshold. Conversely, to retain the equivalent of classical rank $2^{48}$ against the $D = 2^{40}$ adversary with the JNRV20 oracle, Corollary 2 gives $\log_2 r = 48 + 25.9 = 73.9$ (Regime B, since $2 \cdot (48 - 3.4) = 89.2 > 58.6$); for $2^{60}$ it gives 85.9. Under $D = 2^{64}$ the corresponding ranks are $2^{89.2}$ (Regime A) and $2^{109.9}$ (Regime B, since $2 \cdot (60 - 3.4) = 113.2 > 106.6$). Naive "doubling the bits" would have demanded $2^{96}$ and $2^{120}$.

The cost model so far charges every iteration $g$ gates and $d$ depth whichever quantum strategy is used; the next section examines what the two concrete strategies add to that and which one an adversary should choose.

# State Preparation versus Coherent Indexing

This section compares the two ways of injecting the posteriors into a quantum search, proves that block search is never worse in query count up to a constant, and reduces the comparison to the per-iteration overhead $\kappa$ against a measurable property of the instance.

## Two Strategies

### Product-state amplitude amplification (PS-AA).

Under Assumption 1 the prior is a product, so the state

$$\left|\psi_{\mathcal{D}}\rangle = \otimes_{i=1}^{m} \sum_{j\in[n]} \sqrt{p_{i,j}}\right| j\rangle$$

has amplitude $\sqrt{P(k)}$ on every key. Amplitude amplification [11] with $A|0\rangle = |\psi_{\mathcal{D}}\rangle$ and the cipher oracle finds $k^*$ after $\lceil \pi/4\ p_t^{-1/2} \rceil$ iterations if $p_t$ is known, each iteration applying $O_{k^*}$ once and the reflection $AS_0A^{-1}$ once, i.e. two state preparations per iteration.

Algorithm 1: Grover over rank blocks of geometrically increasing size, each block addressed through a coherent map $\iota_s$ from indices to keys. Martin et al. [13] realize $\iota_s$ by weight slices rather than exact rank

blocks (the block is $\{k: W(k) \in [W_1, W_2)\}$ and the map is their getKey), which changes constants only. Each iteration applies $O_{k^*}$, $\iota_s$ and $\iota_s^{-1}$ once, and a reflection about the uniform state on the block.

Both strategies call the cipher oracle once per iteration. The difference is the *number* of iterations—$\approx \pi/4/\sqrt{p_t}$ against $\approx \pi/4\sqrt{r}$ up to the block-ratio constant—and the *per-iteration overhead*: two preparations of $|\psi_{\mathcal{D}}\rangle$ for PS-AA, two evaluations of $\iota_s$ for BS. We write $g_{\text{AA}} = g + 2g_{\text{prep}}$, $g_{\text{BS}} = g + 2g_\iota$ and $\kappa := g_{\text{BS}}/g_{\text{AA}}$; $\kappa$ may be smaller or larger than 1.

## Block Search Never Loses Asymptotically

**Lemma 2 (Rank–probability bound).** For every $k \in \mathcal{K}$, $\text{rank}(k) \cdot P(k) \leq 1$. In particular $\sqrt{r} \leq 1/\sqrt{p_t}$, with equality if and only if the $r$ most probable keys all have probability exactly $p_t$ and no other key has positive probability.

**Proposition 3 (Which strategy wins).**

1. (a) *(Per instance.)* Up to constants, BS uses at most as many oracle queries as PS-AA on every key, by Lemma 2. With the constants of Theorem 1(b) for BS ($c_{\text{BS}} \leq 2.69$ with ratio-2 blocks and unknown rank) and $\pi/4$ for PS-AA with known $p_t$, BS is cheaper in gates if and only if

$$\kappa \cdot \frac{4c_{\text{BS}}}{\pi} < \frac{1}{\sqrt{r\,p_t}} =: \sigma(k^*), \qquad \text{i.e.} \qquad \log_2\kappa < \log_2\sigma(k^*) - 1.77 \text{ bits},$$

   and $\sigma(k^*) \geq 1$ always. We call $\sigma$ the *advice sharpness* of the instance: it equals 1 when the true key sits on a flat plateau of the prior and grows when the true key is far down a tail dominated by a few heavy candidates.

2. (b) *(Average case.)* The expected number of PS-AA iterations over $K \sim \mathcal{D}$ is exactly

$$\mathbb{E}\left[p_K^{-1/2}\right] = \sum_{k\in\mathcal{K}} \sqrt{P(k)} = \prod_{i=1}^{m} \sum_{j\in[n]} \sqrt{p_{i,j}} = 2^{H_{1/2}(\mathcal{D})/2},$$

   whereas BS achieves $\Theta\left(\mathbb{E}\left[\sqrt{G}\right]\right) = 2^{H_{2/3}(\mathcal{D})/2+O(\log m)}$. Since $H_{2/3} \leq H_{1/2}$, PS-AA is exactly quadratic on average relative to classical guesswork $2^{H_{1/2}(\mathcal{D})}$, and the entire super-quadratic average-case gain of Proposition 1 is the gain of block ordering over product-state amplification.

Part (b) gives the evaluator a closed form that needs neither enumeration nor estimation: $\sum_i \log_2 \sum_j \sqrt{p_{i,j}}$ is computed from the $m \times n$ posterior table in $O(mn)$ time and equals $\log_2$ of the expected PS-AA iteration count. Proofs in Appendix A.7.

**Lemma 3 (Product-state preparation).** $|\psi_{\mathcal{D}}\rangle$ is prepared from $|0\rangle^{\otimes m\log_2 n}$ by $m$ independent $(\log_2 n)$-qubit preparations run in parallel. Each uses $n-1$ controlled single-qubit rotations arranged as a binary tree of depth $\log_2 n$ with $2^\ell$ rotations at level $\ell$, i.e. 255 rotations per byte for $n = 256$ [41]; no QRAM is needed. If a rotation to precision $\varepsilon$ costs $t(\varepsilon)$ $T$ gates and depth $\delta(\varepsilon)$, then one preparation costs $m(n-1)\,t(\varepsilon)$ $T$ gates and, with the $2^\ell$ rotations of a level applied as multiplexed rotations, depth $O\big(n\,\delta(\varepsilon)\big)$.

For $m = 16$, $n = 256$ and the ancilla-free Clifford+$T$ synthesis cost $t(\varepsilon) \approx 3\log_2(1/\varepsilon) + O\big(\log\log(1/\varepsilon)\big)$ $T$ gates per $z$-rotation [42], precision $\varepsilon = 2^{-10}$ gives about 4080 rotations and roughly $1.2 \times 10^5$ $T$ gates per preparation *(derived)*, i.e. about $2.4 \times 10^5$ $T$ gates per PS-AA iteration—four to five times the 54 908 $T$ gates of the AES-128 oracle itself [17, Table 8]. PS-AA is therefore QRAM-free but not cheap:

$g_{\mathrm{AA}} \approx g + 2g_{\mathrm{prep}}$ with $g_{\mathrm{prep}}$ comparable to or larger than $g$ in $T$-count. Lower precision is admissible because the posteriors are themselves noisy; the required $\varepsilon$ as a function of the leakage SNR is left open.

**Lemma 4 (Coherent indexing, symbolic).** The map $\iota_s$ of [13] evaluates getKey on a superposition of indices in $O(m^2 n \log n)$ classical operations per call, reading the path-counting array $b$ of $m \times W_{\max}$ integers; realized coherently this costs $g_\iota$ gates and $d_\iota$ depth, with $g_\iota$ dominated by $O(mn)$ comparisons of $(m\log n)$-bit integers against $b$ held in QRAM or in a hard-wired circuit.

We leave $g_\iota$ and $d_\iota$ symbolic here; their gate-level values for $m = 16$, $n = 256$ are the subject of the companion paper [22]. What this paper establishes is the *decision rule*: BS should be used whenever $\log_2\kappa < \log_2\sigma(k^*) - 1.77$, PS-AA otherwise, and Sect. 7 measures the distribution of $\sigma$ on simulated posteriors (median 2.15 bits) so that the rule can be applied once $\kappa$ is known.

**Remark 4 (Hybrid).** The two strategies compose: inside a weight slice one may start from the product state restricted to the slice rather than the uniform state. This keeps the $\sqrt{r}$ scaling of BS while recovering part of the PS-AA gain on plateaus; we do not analyze it further.

With the theory complete, Sect. 7 measures the quantities the theory leaves to data: the rank-versus-traces relation, the trace discount, and the distribution of the advice sharpness $\sigma$.

## Experiments

This section measures the quantities the theory leaves to data: how the rank falls with the number of traces, how many traces a quantum budget is worth, how sharp real posteriors are, and how the four search strategies behave on a toy cipher.

We simulate an unprotected first-round AES-128 S-box leakage $L_{t,i} = \mathrm{HW}\left(S(x_{t,i} \oplus k_i)\right) + \mathcal{N}(0,\sigma^2)$ for trace $t$ and byte $i$, with uniform plaintext bytes and $\sigma \in \{1,2,4,8\}$, i.e. $\mathrm{SNR} = \mathrm{Var(HW)}/\sigma^2 = 2/\sigma^2 \in \{2, 0.5, 0.125, 0.031\}$. Two attacks are run per byte. The CPA attack [23] uses the Hamming-weight model; since the leakage is Gaussian with known $\sigma$, we map scores to posteriors by the exact likelihood $p_{i,j} \propto \exp(-\mathrm{SSE}_{i,j}/2\sigma^2)$, where $\mathrm{SSE}_{i,j}$ is the squared error of hypothesis $j$ over the $q$ traces; this is the Bayesian posterior under a uniform key prior and is monotone in the Pearson correlation. The Gaussian template attack [24] builds one template per S-box output value from $20\,000$ profiling traces on an independent key and returns likelihoods directly. The full-key rank is computed by histogram convolution [5; 3] with $2048$ bins per byte; we record the lower and upper bounds and use their midpoint. Each $(\mathrm{attack}, \sigma, q)$ cell holds $100$ random keys with derived seeds (Appendix B); $q$ ranges over $\{5, 10, 20, \ldots, 1280\}$ and, for $\sigma = 4$, a denser grid of $18$ values up to $240$.

A 4-round substitution–permutation network with the PRESENT S-box on four nibbles, a PRESENT-style bit permutation, a 16-bit key with round keys obtained by rotation ($m = 4$, $n = 16$). Leakage ($\sigma = 1.5$, $q = 8$) and CPA posteriors are generated as above. The quantum runs use a phase-flip oracle on the known key so that success probabilities are computed exactly on a $2^{16}$-dimensional state vector; the full reversible cipher oracle is synthesized separately in Qiskit and only counted. The separation follows standard practice for Grover resource estimation [17].

(U) Grover from the uniform state over all $2^{16}$ keys; (PS) amplitude amplification from $|\psi_{\mathcal{D}}\rangle$ with the iteration count set by the true $p_t$; (BS) block search, Algorithm 1, with exact rank blocks (indexing is free on a simulator); (CL) classical optimal-order enumeration, success probability $\min(1, j/r)$ after $j$ tests. For each of 120 random keys we record the number of oracle calls at which each strategy first reaches success probability $3/4$.

## Rank versus Traces and the Linear Model

Figure 1 shows the median and interquartile range of $\log_2 r(q)$ and the per-instance quantum counterpart $1/2 \log_2 r(q)$ for $\sigma = 4$. The decay is close to linear in $q$ in the middle of the range and flattens at both ends: near 128 bits because the rank is capped, and below about 20 bits because the few remaining wrong candidates are the ones hardest to separate. We therefore state the model on the mid-range only.

**Assumption 3 (Linear log-rank decay).** For fixed SNR and $q$ in the range where $20 \leq \log_2 r(q) \leq 110$, the median $\log_2 r(q)$ is approximately affine in $q$: $\log_2 r(q) \approx \lambda - \mu q$ with $\mu > 0$ depending on the SNR.

| attack | $\sigma$ | SNR | $\lambda$ (bits) | $\mu$ (bits/trace) | $\mu$/SNR | $R^2$ (points) |
|---|---|---|---|---|---|---|
| CPA | 2 | 0.5 | 120.4 | 4.09 | 8.2 | 1.000 (3) |
| CPA | 4 | 0.125 | 118.3 | 1.10 | 8.8 | 0.995 (7) |
| CPA | 8 | 0.031 | 114.7 | 0.263 | 8.4 | 0.999 (3) |
| template | 2 | 0.5 | 120.9 | 4.11 | 8.2 | 1.000 (3) |
| template | 4 | 0.125 | 118.1 | 1.09 | 8.7 | 0.996 (7) |
| template | 8 | 0.031 | 116.9 | 0.259 | 8.3 | 1.000 (3) |

***Table 4.*** *Fitted parameters of Assumption 3 on the median $log_2 r(q)$ over cells with $20 \leq log_2 r \leq 110$ bits. The takeaway is that the decay rate scales as $\mu \approx 8.5\, SNR$ bits per trace across a $16 \times$ range of SNR, and that CPA and templates agree to within 3%.*

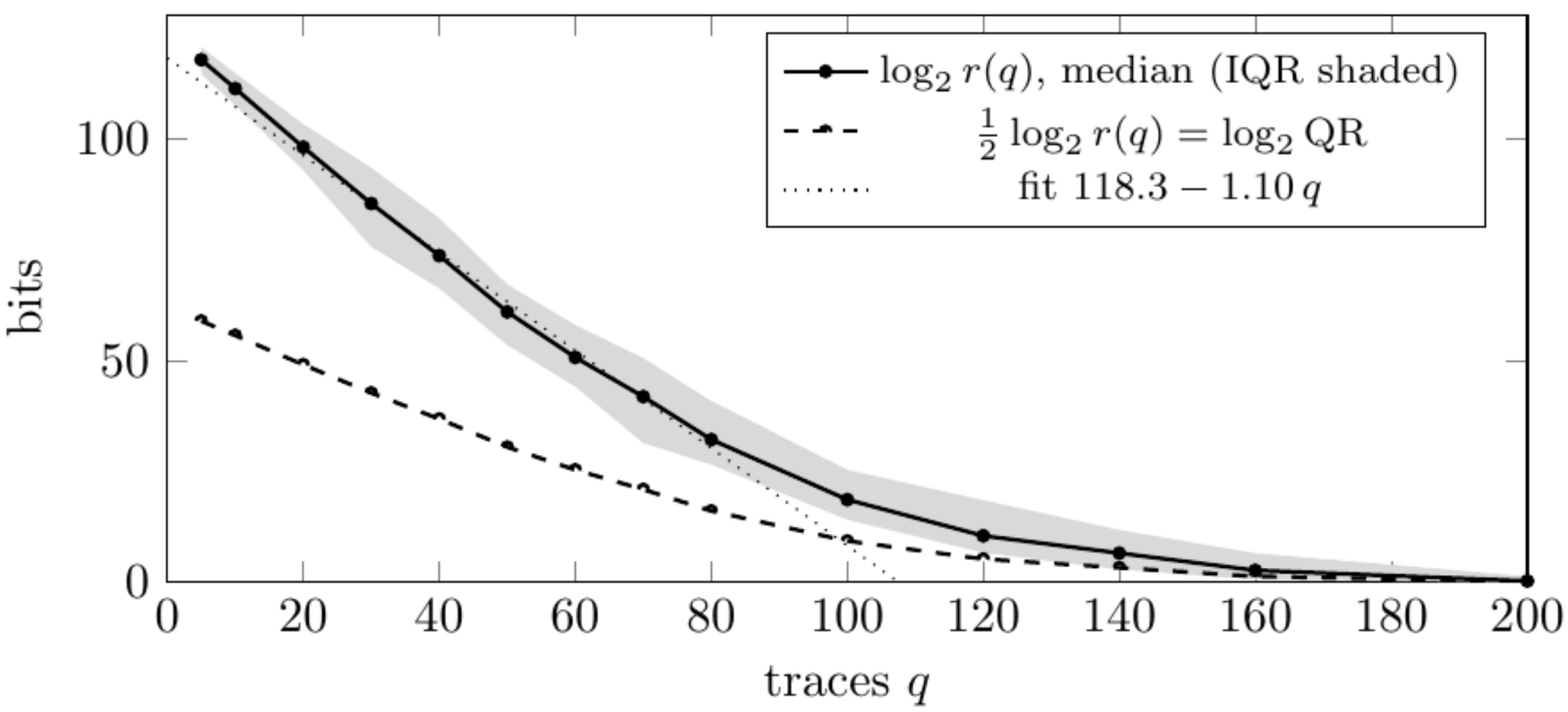

***Table 5.*** *Key rank versus traces for CPA at $SNR = 0.125$ ($\sigma = 4$), 100 keys per point. The takeaway is that $QR$ halves the vertical axis and nothing else; the linear model holds between 20 and 110 bits and flattens outside.*

The decay rate per trace is proportional to the SNR, $\mu \approx$ (8.2–8.8) SNR bits per trace for sixteen bytes (Table 4), which is the expected behavior of a Gaussian likelihood accumulating $\Theta$(SNR) bits of evidence per trace. The discretization factor of Proposition 2 is small: the histogram bounds differ by a median of 0.009 bits and by at most 1.6 bits, the latter only at ranks below $2^4$.

Let the attacker's gate budget be $B$. Define $q_{\mathrm{cl}}(B) = \min\{q: C_{\mathrm{cl}}(r(q)) \le B\}$ and $q_q(B,D) = \min\{q: C_q^D(r(q)) \le B\}$, and the *trace discount* $\Delta q(B,D) = q_{\mathrm{cl}}(B) - q_q(B,D)$: the number of traces the classical attacker must collect beyond what the quantum attacker needs to reach the same budget.

**Proposition 4 (Trace discount under the linear model).** Under Assumption 3 with parameters $(\lambda,\mu)$, and writing $\beta = \log_2 B$,

$$\Delta q(B,D) \;\approx\; \begin{cases} \dfrac{1}{\mu}\left((\beta - \log_2 c_{\mathrm{cl}}) - 2\log_2 \dfrac{\pi g}{4c_{\mathrm{cl}}}\right) & \text{if } B \le B^*, \\ \dfrac{1}{\mu}\log_2 \gamma(D) & \text{if } B > B^*, \end{cases}$$

where $B^* := C_q^D(r_{\mathrm{hi}}) = gD/d$. In Regime A the discount grows by $1/\mu$ traces per doubling of the budget; in Regime B it is a constant equal to $\log_2\gamma$ bits' worth of traces.

*Proof.* The classical attacker needs $\log_2 r = \beta - \log_2 c_{\mathrm{cl}}$; by Corollary 2 the quantum attacker needs $\log_2 r = 2(\beta - \log_2 c_{\mathrm{cl}}) - 2\log_2(\pi g/4c_{\mathrm{cl}})$ in Regime A and $\log_2 r = \beta - \log_2 c_{\mathrm{cl}} + \log_2\gamma$ in Regime B. Invert $\log_2 r(q) = \lambda - \mu q$ and subtract. The boundary $B^*$ is $C_q^D$ at $r_{\mathrm{hi}}$, Lemma 1.

Figure 2 plots the discount measured on the median rank curve for CPA at SNR $= 0.125$ against the prediction of Proposition 4 with the fitted $\mu = 1.10$ and the JNRV20 Maximov oracle, for which $B^* = 2^{71.7}$ at $D = 2^{64}$ and $2^{47.7}$ at $D = 2^{40}$ *(derived)*. Three features of the data are the ones the theory predicts. First, for $D = 2^{40}$ the measured discount rises to 27 traces at $\beta = 50$ and then stays between 20 and 22 traces for $\beta \in [55,105]$; the predicted plateau is $\log_2\gamma(2^{40})/\mu = 25.9/1.10 = 23.5$ traces. Second, for $D = 2^{64}$ the discount keeps growing until $\beta \approx 75$, where it reaches 39–42 traces, and the predicted knee is at $\beta^* = 71.7$ with plateau 45.4. Third, the two depth budgets coincide below $\beta \approx 45$, where neither attack is depth-limited. The prediction overestimates the Regime-A slope below $\beta = 45$ by up to six traces, because there the quantum attacker needs ranks below 20 bits, outside the range of Assumption 3; and above $\beta \approx 100$ every discount falls to zero because the classical attacker already needs fewer than 20 traces, a cap the linear model does not contain. Within the window the measured and predicted discounts agree to within 10%.

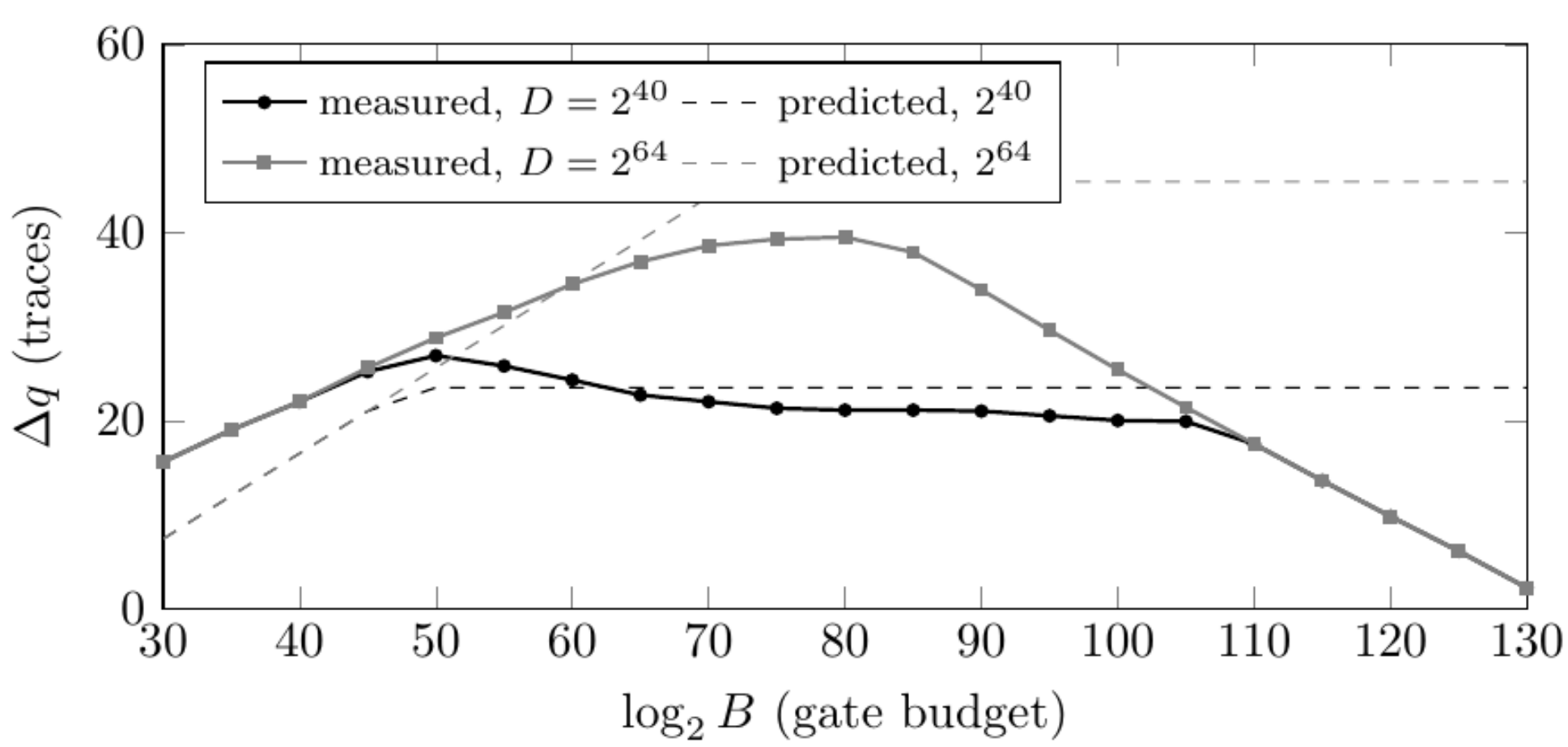

***Figure 2.*** *Trace discount versus gate budget, CPA at SNR* $= 0.125$*, JNRV20 Maximov oracle,* $c_{cl} = 2^{15}$*. The takeaway is the plateau for* $D = 2^{40}$ *from* $\beta \approx 50$ *onward at* 21–22 *traces against a predicted* 23.5*: beyond* $B^*$ *a larger budget buys the depth-limited quantum attacker no further traces, while the* $D = 2^{64}$ *attacker keeps gaining until its own knee near* $\beta = 72$.

## Advice Sharpness on Simulated Posteriors

For each key we record $\sqrt{r}$ and $1/\sqrt{p_t}$ and their ratio $\sigma(k^*) = 1/\sqrt{rp_t}$, the advice sharpness of Proposition 3. Figure 3 is the scatter and Table 5 gives quantiles of $\log_2\sigma$ over the cells whose median rank lies between $2^{20}$ and $2^{100}$, the range an evaluator cares about. The sharpness is remarkably stable: across both attacks, four SNRs and all trace counts, its median is 2.1–2.3 bits, its 90th percentile 2.5–3.3 bits, its 99th percentile below 5.3 bits, and its maximum over 4 000 instances 6.0 bits. In words: on Hamming-weight posteriors the true key's probability is typically $2^{4.3}$ times smaller than $1/\text{rank}$, and never more than $2^{12}$ times smaller.

***Figure 3.*** *Quantiles of* $log_2\sigma(k^*) = log_2\left(1/\sqrt{rp_t}\right)$ *over* 100 *keys per cell, CPA (template values agree within* 0.1 *bit). The takeaway is that the sharpness is almost independent of SNR and trace count: block search can afford about* $2^{2.1}$ *of indexing overhead on the median device and* $2^3$ *on nine devices in ten before product-state amplification becomes cheaper.*

| SNR | $q$ | median $\log_2 r$ | median | 90% | 99% |
|---|---|---|---|---|---|
| 2 | 5 | 62.7 | 1.96 | 2.49 | 4.14 |
| 0.5 | 10 | 79.1 | 2.14 | 2.72 | 3.32 |
| 0.5 | 20 | 38.8 | 2.11 | 2.97 | 4.12 |
| 0.125 | 20 | 98.2 | 2.06 | 2.86 | 3.78 |
| 0.125 | 40 | 73.7 | 2.26 | 2.97 | 4.90 |
| 0.125 | 80 | 32.2 | 2.09 | 2.52 | 3.01 |
| 0.031 | 80 | 94.3 | 2.08 | 3.10 | 4.20 |
| 0.031 | 160 | 71.5 | 2.23 | 2.92 | 3.70 |
| 0.031 | 320 | 30.7 | 2.17 | 3.11 | 4.88 |
| all cells | | | 2.15 | 2.96 | 4.25 |

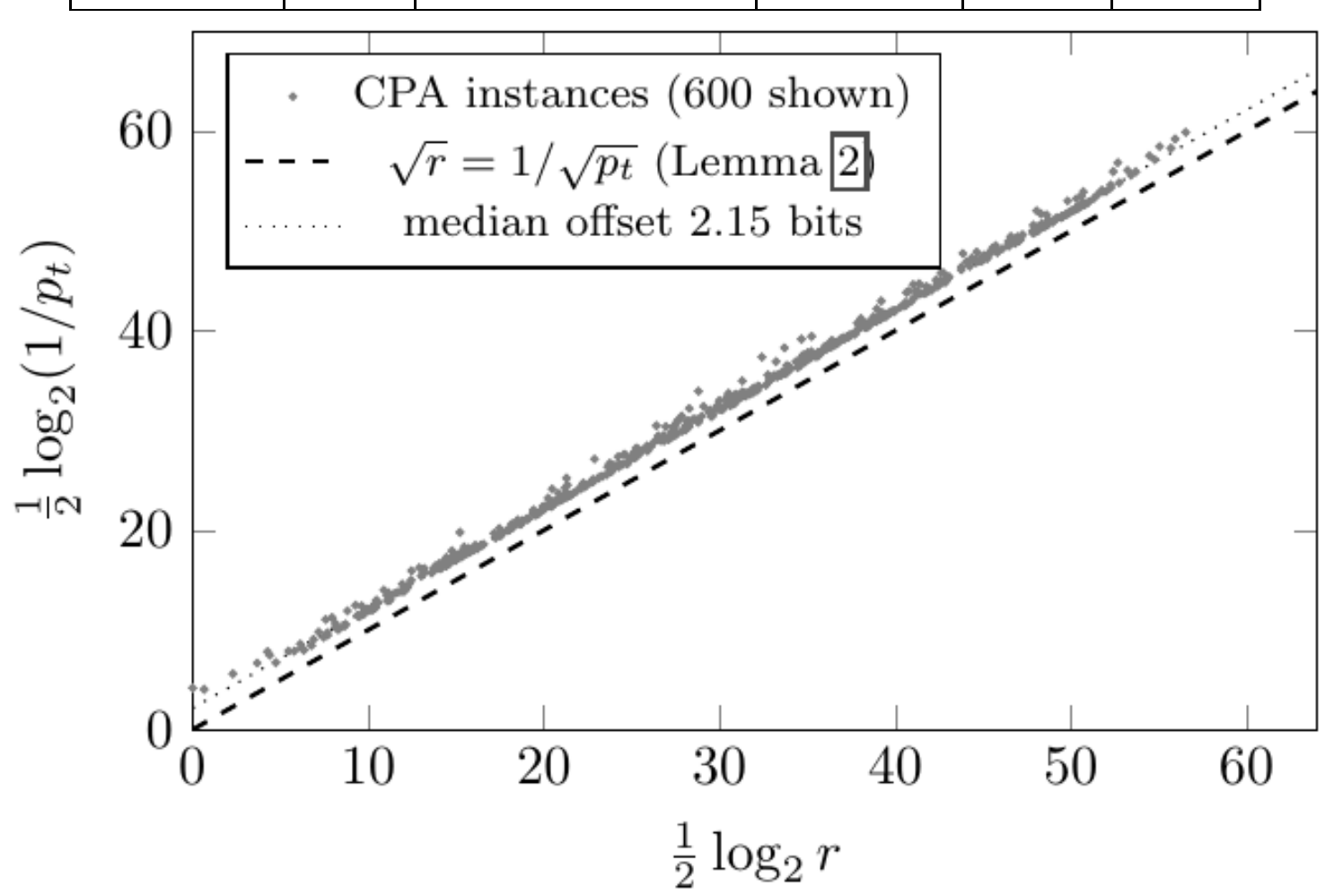

***Table 6.*** *Per-key query cost of block search (*$\sqrt{r}$*) against product-state amplification (*$1/\sqrt{p_t}$*) on CPA posteriors. Every point lies above the diagonal by Lemma 2; the vertical distance is* $log_2\sigma$*, concentrated between* 2 *and* 3 *bits at every rank.*

The constants of Proposition 3 now have values. With ratio-2 blocks and unknown rank, block search needs up to $2.69\sqrt{r}$ calls (Theorem 1(b)); product-state amplification with known $p_t$ needs $\pi/4\, p_t^{-1/2} = \pi/4\, \sigma\sqrt{r}$ calls to reach success near 1 and $\pi/6\, \sigma\sqrt{r}$ to reach $3/4$. Block search is therefore cheaper in queries iff $\log_2\sigma > \log_2(2.69 \cdot 4/\pi) = 1.77$ bits (target probability 1) or $> 2.36$ bits (target $3/4$). Against Table 5, block search wins on about $65\%$ of AES instances at the first threshold and about $40\%$ at the second, *before* any indexing overhead $\kappa$ is charged; every bit of $\log_2\kappa$ moves the threshold up by one bit and removes a further quarter to a third of the instances. The two strategies are thus within a factor 2–4 of each other on realistic posteriors, and the gate-level value of $\kappa$, not the query count, decides between them.

## Quantum Simulation on the Toy Cipher

Figure 4 shows one representative key (rank $329$, $p_t = 3.5 \cdot 10^{-4}$); Table 6 summarizes all $120$ keys. The observations are the ones the theory predicts, including the one the previous paragraph warns about.

Three points deserve comment. First, the uniform strategy reaches $3/4$ at $134$ calls on every key, as it must, which on the toy is *more* than classical enumeration needs for $63$ of the $120$ keys: a $16$-bit key space is too small for Grover to help, and the posteriors are what make the quantum attack worthwhile. Second, block search obeys the bound of Theorem 1(b) on every key with median constant $2.63$, within $2\%$ of the worst-case $2.69$; the block structure makes its success curve a staircase, flat until the block holding the key is reached. Third, product-state amplification is cheaper than block search on $105$ of $120$ keys. This does not contradict Lemma 2, which bounds $1/\sqrt{p_t}$ below by $\sqrt{r}$ without constants: on the toy the median sharpness is $1.35$ bits, below the $1.77$-bit threshold derived above, so the constant-factor penalty of block search for not knowing the rank outweighs the sharpness gain. Among the $22$ keys with sharpness above $1.77$ bits, block search wins on half. On AES posteriors (Table 5), whose sharpness is a bit higher, the balance tips the other way, but only by a factor of about two.
The full reversible oracle for the toy cipher, synthesized in Qiskit with out-of-place lookup-table S-boxes (one 4-controlled $X$ per set output bit, 32 per S-box), a $16$-controlled comparison, and full uncomputation, has width $97$ qubits, $59\,137$ Clifford+$T$ gates ($21\,454$ CNOT, $26\,502$ $T$), and full depth $47\,779$ after transpilation with no ancilla-assisted multi-controlled gates *(measured; a naive design, reported only to place the toy in the cost model)*. Its own quantum window has $\log_2 r_{\mathrm{hi}} = 2(\log_2 D - 15.5 + 0.35) \geq 49$ for $D = 2^{40}$, far beyond a 16-bit key: every search on the toy is in Regime A, as the simulation assumes.

The experiments quantify what the theory parameterizes; Sect. 8 turns both into a rule an evaluation laboratory can apply.

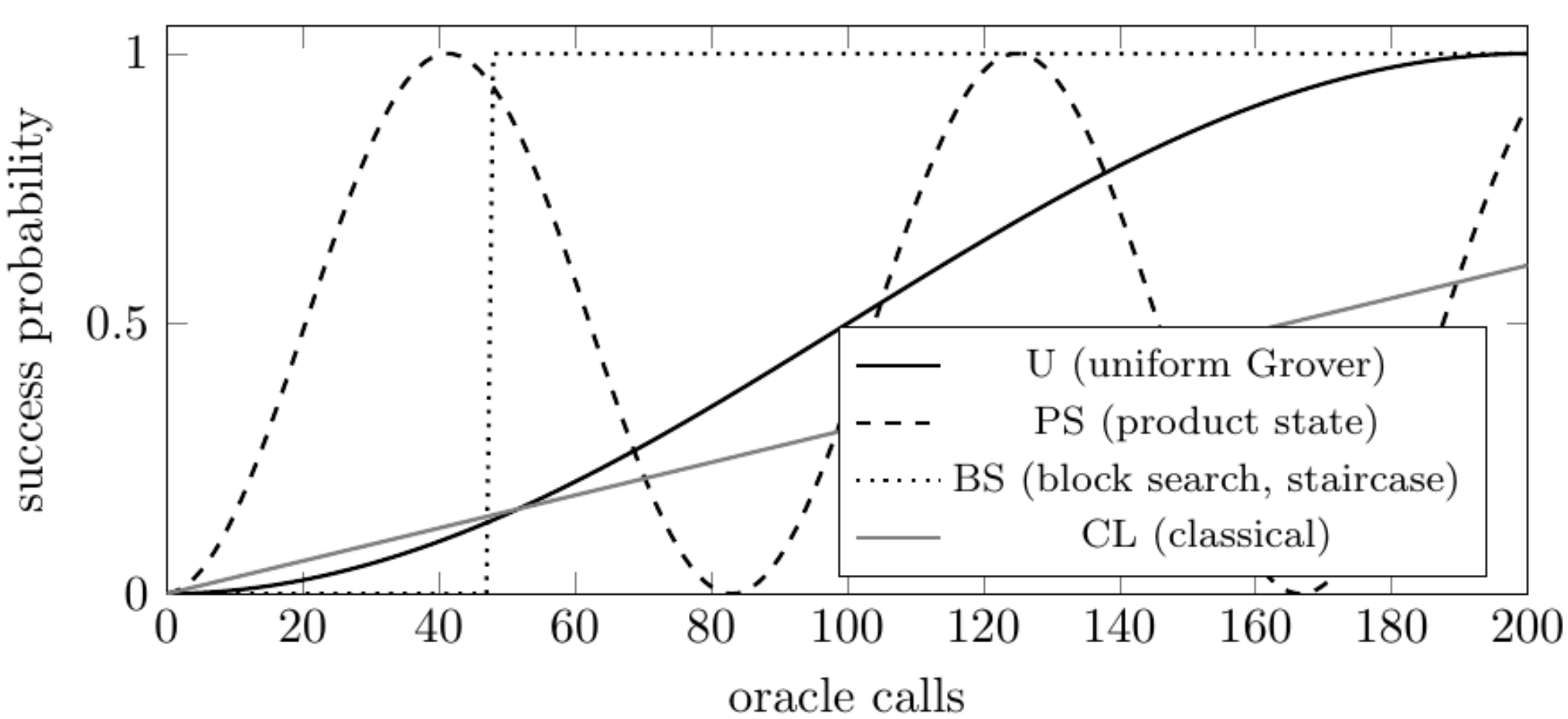

***Figure 4.*** *Toy cipher,* 120 *keys,* $\sigma = 1.5$*,* $q = 8$*: oracle calls to first reach success probability* $3/4$*, normalized by* $\sqrt{r}$ *where meaningful. The takeaway is that block search tracks its* $2.69\sqrt{r}$ *bound and never exceeds it, while product-state amplification is cheaper here because the toy posteriors are flatter (median sharpness* 1.35 *bits) than the AES ones.*

| quantity | median | IQR / range | note |
|---|---|---|---|
| rank $r$ | 74 | 19–314 | max 11 691 |
| $\log_2\sigma$ (bits) | 1.35 | 90%: 1.92 | max 3.77 |
| U: calls to $3/4$ | 134 | 134–134 | $\pi/6 \cdot 2^8 = 134$, independent of posteriors |
| PS: calls to $3/4$, $/\sqrt{r}$ | 1.31 | — | predicted $\pi/6\,\sigma = 1.33$ at median $\sigma$ |
| BS: calls to $3/4$, $/\sqrt{r}$ | 2.63 | — | bound 2.69; satisfied on $120/120$ keys |
| CL: calls to $3/4$ | 37 | — | $= \lceil 0.75\,r \rceil$; 24 keys need $> 300$ |
| BS $\leq$ PS | $15/120$ | | $11/22$ among keys with $\log_2\sigma > 1.77$ |
| PS peak success | 1.00 | $\geq 0.9$ on $119/120$ | min 0.43 (overshoot at $r \leq 2$) |

***Table 7.*** *Success probability versus oracle calls on the 16-bit toy cipher for a key of rank* 329 *and* $p_t = 3.5 \cdot 10^{-4}$*. The takeaway is the ordering PS < BS < U in calls on this flat-posterior instance, with PS oscillating after its peak at* $\lceil \pi/4\, p_t^{-1/2} \rceil = 42$ *calls and BS rising in one step when its block of ranks* $[256{,}512)$ *is searched.*

## Implications for Evaluation Practice

This section states the evaluation rule the preceding results support and maps it onto the frameworks evaluation laboratories currently use.

An evaluator who today reports a classical rank interval $[r_\ell, r_u]$ and compares it with a threshold $r_{\text{th}}$ should additionally report, for each $D \in \{2^{40}, 2^{64}, 2^{96}\}$ and for a named oracle circuit, the interval $[\text{QKR}_D(r_\ell), \text{QKR}_D(r_u)]$ of Corollary 2 and compare it with the *same* $r_{\text{th}}$. Equivalently, the classical rank a device must have to retain the equivalent of $r_{\text{th}}$ against the depth-$D$ quantum adversary is the inverse map of Corollary 2; Table 7 evaluates it for the thresholds in current use. Because every post-2020 AES-128 circuit in Table 3 gives a window boundary within one bit of $2^{59}$ at $D = 2^{40}$ and a constant gain within one bit of $2^{26}$, the rule is insensitive to the choice among modern circuits; only the Grassl-era anchor behind NIST's $2^{170}$ figure differs, by 13 bits.

*Classical rank (bits) required to retain the equivalent of a classical threshold against a depth-limited quantum adversary,* $c_{cl} = 2^{15}$*, derived from Corollary 2; regime in parentheses. Rows give the JNRV20 Maximov oracle and, in brackets, the range over the four post-2020 circuits of Table 3. The takeaway is that the required margin is* $2\times$ *the threshold minus* 6.8 *bits inside the window and* $log_2\gamma \approx 26$ *bits above it, never the* $2\times$ *of the naive rule.*

| threshold $r_{\text{th}}$ | $D = 2^{40}$ | $D = 2^{64}$ | $D = 2^{96}$ |
|---|---|---|---|
| $2^{48}$ | 73.9 (B) [73.8–74.7] | 89.2 (A) [89.1–90.2] | 89.2 (A) [89.1–90.2] |
| $2^{60}$ | 85.9 (B) [85.8–86.7] | 109.9 (B) [109.8–110.7] | 113.2 (A) [113.1–114.2] |
| $2^{80}$ | 105.9 (B) [105.8–106.7] | 129.9 (B) [129.8–130.7] | 153.2 (A) [153.1–154.2] |

The last row answers the remark that a $2^{80}$ cut-off "reduces to $2^{40}$" under Grover [7]: with a depth budget of $2^{40}$ it does not, because the attack is in Regime B and the quantum adversary pays $2^{80-25.9} = 2^{54.1}$ classical-equivalent gates, not $2^{40}$; without a depth budget ($D = 2^{96}$) it reduces to $2^{43.4}$ classical-equivalent gates, which is the naive figure plus the 3.4-bit oracle overhead. Both statements are reached from one formula. In traces, Fig. 2 says the same thing: at $\mathrm{SNR} = 0.125$ a depth-$2^{40}$ quantum adversary is worth at most 22 extra traces to a classical one, however large its gate budget.

The JIL rating for smartcards [8] scores an attack by elapsed time, expertise, knowledge of the target, access, equipment and open samples, and it defers the "remaining strength of cryptographic implementations" to a chapter still under discussion. A quantum-aware extension needs only two additions: (i) an equipment class for a fault-tolerant quantum computer parameterized by its depth budget, and (ii) the instruction to score the exploitation phase by $\mathrm{QKR}_D$ rather than by the classical rank. Everything else in the rating is unchanged, because the identification phase (trace collection, CPA) is classical in the Q1 model.

ISO/IEC 17825:2024 [9] is a conformance test with a low-to-medium assurance level and no adversary model beyond first-order leakage detection; it has no place for a rank threshold and therefore none for a quantum one. The appropriate home for the rule above is a laboratory's evaluation methodology, not the conformance test; we do not propose amending 17825 itself.

The rule charges logical gates and ignores quantum error-correction overhead, which multiplies the quantum cost by a factor that depends on the code and the target error rate and that favors the classical adversary. It assumes the independence of Assumption 1 and the oracle costs of a named circuit. Its purpose is to replace a wrong rule of thumb with a right formula, not to predict the year in which the attack becomes practical.

With the practical consequence stated, we close with what the analysis leaves open.

## Limitations and Future Work

Six limitations are specific enough to be worth naming.

*Coherent indexing and constants.* Block search needs $\iota_s$ realized coherently, either with QRAM holding the path-counting array or with a hard-wired circuit; we carry its cost as the symbolic $\kappa$ and state the decision rule, but the gate-level value of $\kappa$ for $m = 16$, $n = 256$ is left to the companion paper [22]. Section 7.4 shows that the decision is close even at $\kappa = 1$: the sharpness of Hamming-weight posteriors is 2–3 bits and the block-search constant for unknown rank costs 1.8 bits of it, so a $\kappa$ of 2–4 makes product-state amplification the practical strategy on most devices, and then its exactly quadratic average-case behavior (Proposition 3(b)) is the relevant one.

*Error correction.* The cost model counts logical gates and depth. Surface-code overheads for the $T$-heavy AES oracle are large and shift $r_{\mathrm{lo}}$ upward by their logarithm; they do not change $r_{\mathrm{hi}}$, which depends only on depth, nor the structure of the two regimes.

*Independence and leakage model.* Masked implementations, and template attacks that exploit key-schedule or MixColumns relations, produce posteriors that are not products; Lemma 2 and Theorem 1 do not need independence, but the product-state preparation of Lemma 3 and the entropy identities do. Our experiments use simulated Hamming-weight leakage with Gaussian noise; the sharpness values of Table 5 are a property of that leakage model and should be re-measured on profiled posteriors from a real device (ASCAD [43] is the natural candidate) before being used as constants.

*Oracle uniqueness.* Assumption 2 with one plaintext–ciphertext pair fails with probability about $r \cdot 2^{-128}$; a second pair doubles $g$ and $d$ and shifts every window boundary by one bit.

*Published circuits.* Table 3 takes $(g, d)$ from the full Grover oracle of [17] and, for the three later circuits, from the encryption circuits as tabulated in [21], with one-qubit Clifford and measurement counts estimated by the JNRV20 ratio; the resulting $g$ values are marked as estimates and affect $r_{\mathrm{lo}}$ and $\gamma$ by less than one bit. A lower $d$ moves $r_{\mathrm{hi}}$ up by $2\log_2$ of the improvement.

*Other ciphers.* The analysis is parameterized by $(g, d, c_{\mathrm{cl}})$ and applies to any block cipher for which these are known; quantum circuits and the corresponding window for the Ukrainian standard Kalyna (DSTU 7624:2014), whose modular key addition and four distinct S-boxes make its oracle cost unlike AES's, are the subject of the companion paper.

Beyond these, the questions we consider most worth pursuing are a tighter constant for block search with unknown rank (Montanaro's $a^s$ blocks with optimized $a$, which the 1.77-bit threshold makes directly valuable), a quantum rank *estimator* based on amplitude estimation over weight slices, and the hybrid of Remark 4.

## Conclusion

For a device whose key has residual rank $r$ after a side-channel attack, the best quantum adversary with classical access to the posteriors needs $\Theta(\sqrt{r})$ cipher evaluations, no fewer; the super-quadratic exponents reported for search with advice are averages over the prior and do not apply to any single device. Under a depth budget $D$, the quadratic advantage holds only for ranks up to $(4D/\pi d)^2$ and collapses to a constant factor $\gamma(D) = 16c_{\mathrm{cl}}D/(\pi^2 gd)$ beyond, so that with every post-2020 AES-128 oracle and $D = 2^{40}$ the window closes within a bit of rank $2^{59}$ and the residual gain is within a bit of $2^{26}$; in traces, at $\mathrm{SNR} = 0.125$, that gain is worth about 22 traces and no more. The quantum-aware key rank $\mathrm{QKR}_D$ packages this into the unit evaluation laboratories already use and is computable from the rank estimators they already run. On Hamming-weight posteriors the two ways of exploiting the advice, block search and product-state amplification, are within a factor of two to four of each other, so the cost of coherent indexing, not the query count, will decide which one a real adversary runs.

# Appendix
# Proofs

## Proof of Theorem 1

### (a) Lower bound.

Fix the posteriors; the ordering of $\mathcal{K}$ and the class $B_r$ are then fixed and known to $\mathcal{A}$. Let $M = |B_r| = \lfloor r/2 \rfloor \geq 4$. Consider the family of $M$ oracles $\{O_k : k \in B_r\}$. For $k \neq k'$ in $B_r$ these are distinct indicator functions on $\mathcal{K}$ differing in exactly two positions, both inside $B_r$. Any algorithm that, for every $k \in B_r$, outputs $k$ with probability $\geq p$ when run with $O_k$, solves unstructured search with a unique marked element among $M$ candidates: compose $\mathcal{A}$ with the fixed injection $B_r \hookrightarrow \mathcal{K}$ and note that queries outside $B_r$ return the same answer for every member of the family, so they can be answered without the oracle. Zalka's theorem [18, Sect. 3] states that after $j$ queries no quantum algorithm attains success probability exceeding $\sin^2\big((2j+1)\theta_M\big)$ with $\sin\theta_M = 1/\sqrt{M}$, uniformly over the marked element; the same bound holds for the success probability averaged over a uniformly random marked element [18; 31]. Requiring $\sin^2\big((2j+1)\theta_M\big) \geq p$ gives $j \geq \big(\arcsin\sqrt{p}/\theta_M - 1\big)/2 \geq c_p\sqrt{M}\,\big(1 - o(1)\big) = c_p\sqrt{r/2}\,\big(1 - o(1)\big)$. For an algorithm with a random number of queries, apply the bound to the algorithm truncated at its expected query count times a constant and use Markov's inequality, losing a constant factor. ▫

### (b) Upper bound.

Let $\mathrm{rank}(k^*) = r$ and $L = \lfloor \log_2 r \rfloor$, so $k^* \in R_L$ (or among the first seven keys if $r \leq 7$). For $s < L$ the block $R_s$ contains no marked key; Grover's algorithm then rotates nothing, the measured candidate is a uniformly random element of $R_s$, and the classical test rejects it. For $s = L$ the block has $2^L$ elements and exactly one marked, and $\lfloor \pi/4\, 2^{L/2} \rfloor$ iterations succeed with probability $\geq 1 - 2^{-L}$ [29]. Summing the query counts for $s = 3, \ldots, L$,

$$\sum_{s=3}^{L} \lfloor \pi/4\, 2^{s/2} \rfloor \;\leq\; \pi/4 \cdot \frac{2^{(L+1)/2}}{\sqrt{2}-1} \;\leq\; \pi/4 \cdot \frac{\sqrt{2}}{\sqrt{2}-1}\sqrt{r} \;<\; 2.69\sqrt{r}\,,$$

plus one classical test per block and seven initial tests, which is the $O(\log r)$ term. If the Grover run on $R_L$ fails, the algorithm continues to larger blocks and never finds $k^*$; the failure probability is at most $2^{-L} \leq 1/8$ for $L \geq 3$, so success is at least $7/8 \geq 3/4$. The algorithm never uses $r$. ▫

## Proof of Proposition 1

Arıkan [35, Theorem 1] proves, for any guessing strategy on a random variable $X$ with distribution $\mathcal{D}$ over a finite set of size $N$ and any $\rho > 0$,

$$(1 + \ln N)^{-\rho} \left(\sum_x \mathcal{D}(x)^{1/(1+\rho)}\right)^{1+\rho} \;\leq\; \mathbb{E}[G^\rho]\,,$$

and that the optimal (decreasing-probability) strategy attains $\mathbb{E}[G^\rho] \leq (\sum_x \mathcal{D}(x)^{1/(1+\rho)})^{1+\rho}$. Since $\log_2(\sum_x \mathcal{D}(x)^\alpha)^{1/(1-\alpha)} = H_\alpha(\mathcal{D})$ with $\alpha = 1/(1+\rho)$, this reads $\log_2 \mathbb{E}[G^\rho] = \rho H_{1/(1+\rho)}(\mathcal{D}) + O(\rho \log\log N)$. Taking $\rho = 1$ gives (1) with $\alpha = 1/2$; taking $\rho = 1/2$ gives (2) with $\alpha = 2/3$; for a product distribution $N = n^m$ and $\log\log N = O(\log m)$, and Rényi entropy is additive over products, $H_\alpha(\mathcal{D}) = \sum_i H_\alpha(\mathcal{D}_i)$. Montanaro [12, Theorem 1] shows that the expected number of queries of search with advice $\mathcal{D}$ is $\Theta\left(\mathbb{E}\left[\sqrt{G}\right]\right)$ and that this is optimal up to a constant. Bashiri et al. [14] draw the same conclusion for the i.i.d. case; the statement above is the non-identical product case, which Arıkan's bound covers verbatim. The inequality $H_{2/3} \leq H_{1/2}$ is monotonicity of Rényi entropy in its order, with equality iff $\mathcal{D}$ is uniform on its support. ▫

## Proof of Lemma 1

If $I_p(r)\, d \leq D$, a single machine runs all $I_p(r)$ iterations within the budget; the cost is $I_p(r) g$. Otherwise, with inner parallelization over $S$ machines each holding a part of size $r/S$ of the block, each machine needs $I_p(r/S) \approx c_p\sqrt{r/S}$ iterations of depth $d$; the constraint $c_p\sqrt{r/S}\, d \leq D$ gives $S \geq c_p^2 r d^2/D^2$ (Jaques et al. 2020, Eq. (8)), and the total gate count is $S \cdot c_p\sqrt{r/S}\, g = c_p\sqrt{rS}\, g = c_p^2\, r\, g\, d/D$ (Jaques et al. 2020, Eq. (9)). The regime boundary is $c_p\sqrt{r}\, d = D$, i.e. $r_{\text{hi}} = \left(D/\left(c_p d\right)\right)^2$, and substituting it into either branch yields $gD/d$. That inner parallelization is the right choice and that the $\sqrt{S}$ loss cannot be avoided is [17, Sect. 2.3] and [18]. ▫

## Proof of Theorem 2

(i) In Regime A, $C_q^D(r) = c_p\sqrt{r}\, g < c_{\text{cl}} r$ iff $\sqrt{r} > c_p g/c_{\text{cl}}$ iff $r > r_{\text{lo}}$. (ii) In Regime B, $C_{\text{cl}}(r)/C_q^D(r) = c_{\text{cl}} r/\left(c_p^2 r g d/D\right) = \gamma(D)$. (iii) $r_{\text{lo}} < r_{\text{hi}}$ iff $c_p g/c_{\text{cl}} < D/\left(c_p d\right)$ iff $c_p^2 g d < c_{\text{cl}} D$ iff $\gamma > 1$, and $r_{\text{hi}}/r_{\text{lo}} = (D c_{\text{cl}}/\left(c_p^2 g d\right))^2 = \gamma^2$. ▫

## Proof of Corollary 2

Divide Lemma 1 by $c_{\text{cl}}$ and substitute $\gamma$. Continuity at $r_{\text{hi}}$: $\pi/4\, g/c_{\text{cl}}\sqrt{r_{\text{hi}}} = gD/dc_{\text{cl}} = r_{\text{hi}}/\gamma$. The inverse map solves each branch for $r$ and keeps the solution consistent with its branch. ▫

## Arithmetic behind Table 3

With $\log_2 c_p = \log_2(\pi/4) = -0.3485$ and $\log_2 c_{\text{cl}} = 15$: $\log_2 r_{\text{lo}} = 2(\log_2 g - 15 - 0.3485)$; $\log_2 r_{\text{hi}} = 2(\log_2 D - \log_2 d + 0.3485)$; $\log_2 \gamma = 15 + \log_2 D + 0.697 - \log_2(gd)$, where $0.697 = \log_2(16/\pi^2)$. For JNRV20 Maximov MC: $g = 447\,986 = 2^{18.773}$, $d = 2086 = 2^{11.027}$, $gd = 2^{29.80}$; hence $\log r_{\text{lo}} = 6.85$, $\log r_{\text{hi}}(2^{40}) = 2 \times 29.32 = 58.64$, $\log\gamma(2^{40}) = 15 + 40 + 0.697 - 29.80 = 25.90$. For JNRV20 in-place MC: $g = 445\,376 = 2^{18.765}$, $d = 2816 = 2^{11.459}$, $gd = 2^{30.22}$. For GLRS16: $g =$

$2^{87.454-63.651} = 2^{23.80}$, $d = 2^{81.214-63.651} = 2^{17.56}$, $gd = 2^{41.37}$, where $\log_2\lfloor\pi/4\, 2^{64}\rfloor = 63.651$, $\log_2 1.37 = 0.454$, $\log_2 1.16 = 0.214$. For LPZW23 only $d$ enters $r_{\text{hi}}$: $d = 730 = 2^{9.51}$ or $1460 = 2^{10.51}$. Sanity check: $c_p^2\, 2^{128}\, gd/2^{40}$ with the JNRV20 Maximov values equals $2^{-0.697+128+29.80-40} = 2^{117.1} = 1.07 \cdot 2^{117}$, matching [17, Table 12].

## Proofs for Section 6

### Lemma 2.

The $\text{rank}(k) - 1$ keys ranked above $k$ each have probability $\geq P(k)$ (ties are broken by a fixed order, so a tied key ranked above has probability exactly $P(k)$). Hence $1 \geq \sum_{k'} P\,(k') \geq \text{rank}(k) \cdot P(k)$. Equality forces every key of rank $\leq \text{rank}(k)$ to have probability $P(k)$ and all others probability 0. ▫

### Proposition 3(a).

PS-AA needs $\lceil \pi/4\, p_t^{-1/2} \rceil$ iterations when $p_t$ is known and at most a constant factor more otherwise (fixed-point search [30] or exponential search over the iteration count [29]); BS needs at most $2.69\sqrt{r}$ iterations by Theorem 1(b). Lemma 2 gives $\sqrt{r} \leq p_t^{-1/2}$. In gates, BS costs at most $c_{\text{BS}}\sqrt{r}\, g_{\text{BS}}$ with $c_{\text{BS}} = \pi/4 \cdot \sqrt{2}/\sqrt{2} - 1 < 2.69$ and PS-AA costs $\pi/4\, p_t^{-1/2} g_{\text{AA}} = \pi/4\, \sigma \sqrt{r}\, g_{\text{AA}}$; the former is smaller iff $\kappa \cdot 4c_{\text{BS}}/\pi < \sigma$, and $4c_{\text{BS}}/\pi = \sqrt{2}/(\sqrt{2} - 1) = 3.41 = 2^{1.77}$. ▫

### Proposition 3(b).

$\mathbb{E}_{K\sim\mathcal{D}}\left[P(K)^{-1/2}\right] = \sum_k P\,(k)\, P(k)^{-1/2} = \sum_k \sqrt{P(k)}$. By definition $H_{1/2}(\mathcal{D}) = 2\log_2 \sum_k \sqrt{P(k)}$, so the sum is $2^{H_{1/2}(\mathcal{D})/2}$; under Assumption 1, $\sum_k \sqrt{P(k)} = \prod_i \sum_j \sqrt{p_{i,j}}$. The BS bound is (2), and $H_{2/3} \leq H_{1/2}$ by monotonicity. ▫

### Lemma 3.

Grover and Rudolph [41]: to prepare $\sum_j \sqrt{p_j}\, |j\rangle$ on $\ell$ qubits, at level $t$ ($0 \leq t < \ell$) apply to qubit $t$, controlled on the first $t$ qubits, the rotation by $\theta_{t,x} = \arccos\sqrt{\Pr[\text{bit}_t = 0 \mid \text{prefix} = x]}$ for each prefix $x \in \{0,1\}^t$; that is $2^t$ controlled rotations at level $t$ and $2^\ell - 1$ in total. The $m$ registers are disjoint, so the $m$ preparations run in parallel. The $2^t$ rotations of one level act on the same target qubit and are therefore sequential as a multiplexed rotation of length $2^t$, giving depth $\sum_t 2^t\, \delta(\varepsilon) = O\big(n\, \delta(\varepsilon)\big)$; $T$-count follows from $t(\varepsilon)$ per rotation. ▫

### Proposition 2.

Monotonicity of $x \mapsto \sqrt{x}$ and of $r \mapsto C_q^D(r)$ (both branches of Lemma 1 are increasing and agree at the boundary) transfers the interval $[r_\ell/\rho, \rho r_u]$ for $\text{rank}(k^*)$ to intervals for QR and $\text{QKR}_D$. The $O(mn)$ computation of $\sum_i H_{2/3}\,(\mathcal{D}_i)$ is immediate from Assumption 1. ▫

## Reproducibility

### Code.

Supplied with the arXiv submission as `code/` (MIT license); a public repository link will be added in v2. Directory layout: `leakage/sim_rank.py` (simulation, CPA, templates, histogram rank), `leakage/analyze.py` (fits, discount, sharpness, figure data), `quantum/toy_sim.py` (exact state-vector runs of U, PS, BS, CL), `quantum/toy_oracle.py` (Qiskit synthesis and gate count of the toy oracle), `window/tables.py` with `circuits.csv` (regenerates Tables 3 and 7), `figures/` (pgfplots data files).

### Software versions.

Python 3.13, NumPy 2.5.3, SciPy 1.18.1, Qiskit 2.5.2 (transpiler only; state vectors are NumPy). No external side-channel library is used.

### Seeds and sizes.

Master seed `20261014`; the key, plaintexts and noise of cell $(\sigma_i, q_j, k)$ are drawn from `default_rng([20261014, i, j, k])`, templates from `[20261014, i, 7]`, toy keys from `[20261014, 99]`. $N_{\text{keys}} = 100$ per $(\sigma, q)$ cell for AES (9 000 instances per attack), 120 keys for the toy cipher; $\sigma \in \{1,2,4,8\}$; $q \in \{5,10,20,40,80,160,320,640,1280\}$ plus $\{30,50,60,70,100,120,140,200,240\}$ for $\sigma = 4$; 2048 histogram bins per byte; 20 000 profiling traces.

### Regenerating the tables.

`python window/tables.py -inputs circuits.csv -ccl 15` prints Tables 3 and 7; the shipped `circuits.csv` contains the rows of Table 2 with their sources.

# References

[1] Veyrat-Charvillon, N., Gérard, B., Renauld, M., & Standaert, F.-X. (2013). An optimal key enumeration algorithm and its application to side-channel attacks. In *Selected Areas in Cryptography – SAC 2012* (LNCS 7707, pp. 390–406). Springer. https://doi.org/10.1007/978-3-642-35999-6_25

[2] Martin, D. P., O'Connell, J. F., Oswald, E., & Stam, M. (2015). Counting keys in parallel after a side channel attack. In *Advances in Cryptology – ASIACRYPT 2015* (LNCS 9453, pp. 313–337). Springer. https://doi.org/10.1007/978-3-662-48800-3_13

[3] Poussier, R., Standaert, F.-X., & Grosso, V. (2016). Simple key enumeration (and rank estimation) using histograms: An integrated approach. In *Cryptographic Hardware and Embedded Systems – CHES 2016* (LNCS 9813, pp. 61–81). Springer. https://doi.org/10.1007/978-3-662-53140-2_4

[4] Veyrat-Charvillon, N., Gérard, B., & Standaert, F.-X. (2013). Security evaluations beyond computing power. In *Advances in Cryptology – EUROCRYPT 2013* (LNCS 7881, pp. 126–141). Springer. https://doi.org/10.1007/978-3-642-38348-9_8

[5] Glowacz, C., Grosso, V., Poussier, R., Schüth, J., & Standaert, F.-X. (2015). Simpler and more efficient rank estimation for side-channel security assessment. In *Fast Software Encryption – FSE 2015* (LNCS 9054, pp. 117–129). Springer. https://doi.org/10.1007/978-3-662-48116-5_6

[6] David, L., & Wool, A. (2019). Poly-logarithmic side channel rank estimation via exponential sampling. In *Topics in Cryptology – CT-RSA 2019* (LNCS 11405, pp. 330–349). Springer. https://doi.org/10.1007/978-3-030-12612-4_17

[7] Oswald, E. (2021). *Side channels: What are they, what can they do, and how should we respond?* [Presentation slides]. NIST Crypto Reading Club. https://csrc.nist.gov/csrc/media/Presentations/2021/crypto-club-2021-side-channels-2/images-media/crclub-2021-side-channels-2.pdf

[8] Joint Interpretation Library. (2024). *Application of attack potential to smartcards and similar devices* (Version 3.2.1). Common Criteria Portal.

[9] ISO/IEC. (2024). *ISO/IEC 17825:2024 — Testing methods for the mitigation of non-invasive attack classes against cryptographic modules* (2nd ed.).

[10] Grover, L. K. (1996). A fast quantum mechanical algorithm for database search. In *Proceedings of the 28th Annual ACM Symposium on Theory of Computing (STOC '96)* (pp. 212–219). ACM. https://doi.org/10.1145/237814.237866

[11] Brassard, G., Høyer, P., Mosca, M., & Tapp, A. (2002). Quantum amplitude amplification and estimation. In *Quantum Computation and Information* (Contemporary Mathematics, Vol. 305, pp. 53–74). American Mathematical Society. https://doi.org/10.1090/conm/305/05215

[12] Montanaro, A. (2011). Quantum search with advice. In *Theory of Quantum Computation, Communication, and Cryptography – TQC 2010* (LNCS 6519, pp. 77–93). Springer. https://doi.org/10.1007/978-3-642-18073-6_7
[13] Martin, D. P., Montanaro, A., Oswald, E., & Shepherd, D. (2018). Quantum key search with side channel advice. In *Selected Areas in Cryptography – SAC 2017* (LNCS 10719, pp. 407–422). Springer. https://doi.org/10.1007/978-3-319-72565-9_21
[14] Bashiri, T., Glaser, T., May, A., & Nowakowski, J. (2026). Super-quadratic quantum speed-ups and guessing many likely keys. In *Advances in Cryptology – EUROCRYPT 2026* (LNCS 16541, pp. 457–480). Springer. https://doi.org/10.1007/978-3-032-25291-3_16
[15] Schubert, M., Paskarbeit, J., Kramer, J., Seifert, J.-P., & Margraf, M. (2026). *Certified quantum speed-up exponents for guessing with side information* (arXiv:2609.28226). arXiv. https://doi.org/10.48550/arXiv.2609.28226
[16] National Institute of Standards and Technology. (2016). *Submission requirements and evaluation criteria for the post-quantum cryptography standardization process*. NIST.
[17] Jaques, S., Naehrig, M., Roetteler, M., & Virdia, F. (2020). Implementing Grover oracles for quantum key search on AES and LowMC. In *Advances in Cryptology – EUROCRYPT 2020* (LNCS 12106, pp. 280–310). Springer. https://doi.org/10.1007/978-3-030-45724-2_10
[18] Zalka, C. (1999). Grover's quantum searching algorithm is optimal. *Physical Review A, 60*(4), 2746–2751. https://doi.org/10.1103/PhysRevA.60.2746
[19] Huang, Z., & Sun, S. (2022). Synthesizing quantum circuits of AES with lower T-depth and less qubits. In *Advances in Cryptology – ASIACRYPT 2022* (LNCS 13793, pp. 614–644). Springer. https://doi.org/10.1007/978-3-031-22969-5_21
[20] Liu, Q., Preneel, B., Zhao, Z., & Wang, M. (2023). Improved quantum circuits for AES: Reducing the depth and the number of qubits. In *Advances in Cryptology – ASIACRYPT 2023* (LNCS 14440). Springer. https://doi.org/10.1007/978-981-99-8727-6_3
[21] Jiang, L.-L., Cai, B.-B., Gao, F., Qin, S.-J., Jin, Z.-P., & Wen, Q.-Y. (2025). Constructing resource-efficient quantum circuits for AES. *Frontiers in Physics, 13*, 1582819. https://doi.org/10.3389/fphy.2025.1582819
[22] Kotukh, Y. (2026). *Quantum resource estimates for Kalyna (DSTU 7624:2014) and the cost of side-channel-assisted key search under depth constraints* [Manuscript in preparation].
[23] Brier, E., Clavier, C., & Olivier, F. (2004). Correlation power analysis with a leakage model. In *Cryptographic Hardware and Embedded Systems – CHES 2004* (LNCS 3156, pp. 16–29). Springer. https://doi.org/10.1007/978-3-540-28632-5_2
[24] Chari, S., Rao, J. R., & Rohatgi, P. (2003). Template attacks. In *Cryptographic Hardware and Embedded Systems – CHES 2002* (LNCS 2523, pp. 13–28). Springer. https://doi.org/10.1007/3-540-36400-5_3
[25] David, L., & Wool, A. (2017). A bounded-space near-optimal key enumeration algorithm for multi-subkey side-channel attacks. In *Topics in Cryptology – CT-RSA 2017* (LNCS 10159, pp. 311–327). Springer. https://doi.org/10.1007/978-3-319-52153-4_18
[26] Martin, D. P., Mather, L., & Oswald, E. (2018). Two sides of the same coin: Counting and enumerating keys post side-channel attacks revisited. In *Topics in Cryptology – CT-RSA 2018* (LNCS 10808, pp. 394–412). Springer. https://doi.org/10.1007/978-3-319-76953-0_21
[27] Bogdanov, A., Kizhvatov, I., Manzoor, K., Tischhauser, E., & Witteman, M. (2016). Fast and memory-efficient key recovery in side-channel attacks. In *Selected Areas in Cryptography – SAC 2015* (LNCS 9566, pp. 310–327). Springer. https://doi.org/10.1007/978-3-319-31301-6_19
[28] Villanueva-Polanco, R. (2019). A comprehensive study of the key enumeration problem. *Entropy, 21*(10), 972. https://doi.org/10.3390/e21100972
[29] Boyer, M., Brassard, G., Høyer, P., & Tapp, A. (1998). Tight bounds on quantum searching. *Fortschritte der Physik, 46*(4–5), 493–505. https://doi.org/10.1002/(SICI)1521-3978(199806)46:4/5<493::AID-PROP493>3.0.CO;2-P

[30] Yoder, T. J., Low, G. H., & Chuang, I. L. (2014). Fixed-point quantum search with an optimal number of queries. *Physical Review Letters, 113*(21), 210501. https://doi.org/10.1103/PhysRevLett.113.210501
[31] Bennett, C. H., Bernstein, E., Brassard, G., & Vazirani, U. (1997). Strengths and weaknesses of quantum computing. *SIAM Journal on Computing, 26*(5), 1510–1523. https://doi.org/10.1137/S0097539796300933
[32] Banegas, G., & Villanueva-Polanco, R. (2023). On searching for keys in quantum cold boot attacks. *Cryptography and Communications, 15*, 495–523. https://doi.org/10.1007/s12095-022-00625-z
[33] Albrecht, M. R., & Shen, Y. (2022). *Quantum augmented dual attack* (Cryptology ePrint Archive, Report 2022/656). IACR.
[34] Høyer, P., Mosca, M., & de Wolf, R. (2003). Quantum search on bounded-error inputs. In *Automata, Languages and Programming – ICALP 2003* (LNCS 2719, pp. 291–299). Springer. https://doi.org/10.1007/3-540-45061-0_25
[35] Arıkan, E. (1996). An inequality on guessing and its application to sequential decoding. *IEEE Transactions on Information Theory, 42*(1), 99–105. https://doi.org/10.1109/18.481781
[36] Zou, J., Wei, Z., Sun, S., Liu, X., & Wu, W. (2020). Quantum circuit implementations of AES with fewer qubits. In *Advances in Cryptology – ASIACRYPT 2020* (LNCS 12492, pp. 697–726). Springer. https://doi.org/10.1007/978-3-030-64834-3_24
[37] Jang, K., Baksi, A., Kim, H., Song, G., Seo, H., & Chattopadhyay, A. (2025). Quantum analysis of AES. *IACR Communications in Cryptology, 2*(1). https://doi.org/10.62056/ay11zo-3y
[38] Grassl, M., Langenberg, B., Roetteler, M., & Steinwandt, R. (2016). Applying Grover's algorithm to AES: Quantum resource estimates. In *Post-Quantum Cryptography – PQCrypto 2016* (LNCS 9606, pp. 29–43). Springer. https://doi.org/10.1007/978-3-319-29360-8_3
[39] Bonnetain, X., Naya-Plasencia, M., & Schrottenloher, A. (2019). Quantum security analysis of AES. *IACR Transactions on Symmetric Cryptology, 2019*(2), 55–93. https://doi.org/10.13154/tosc.v2019.i2.55-93
[40] Kaplan, M., Leurent, G., Leverrier, A., & Naya-Plasencia, M. (2016). Breaking symmetric cryptosystems using quantum period finding. In *Advances in Cryptology – CRYPTO 2016* (LNCS 9815, pp. 207–237). Springer. https://doi.org/10.1007/978-3-662-53008-5_8
[41] Grover, L., & Rudolph, T. (2002). *Creating superpositions that correspond to efficiently integrable probability distributions* (arXiv:quant-ph/0208112). arXiv. https://doi.org/10.48550/arXiv.quant-ph/0208112
[42] Ross, N. J., & Selinger, P. (2016). Optimal ancilla-free Clifford+T approximation of z-rotations. *Quantum Information and Computation, 16*(11–12), 901–953. https://doi.org/10.26421/QIC16.11-12-1
[43] Benadjila, R., Prouff, E., Strullu, R., Cagli, E., & Dumas, C. (2020). Deep learning for side-channel analysis and introduction to ASCAD database. *Journal of Cryptographic Engineering, 10*(2), 163–188. https://doi.org/10.1007/s13389-019-00220-8